%% file: HIGG-2017-04-PAPER.tex
\pdfoutput=1
\newcommand*{\ATLASLATEXPATH}{latex/}
\documentclass[cernpreprint, atlasdraft=true, texlive=2016, UKenglish, LANGEDIT=true, LANGSHOW=false]{\ATLASLATEXPATH atlasdoc}

\usepackage{\ATLASLATEXPATH atlaspackage}
\usepackage{\ATLASLATEXPATH atlasbiblatex}
\usepackage{amsmath,amssymb,latexsym} 
\usepackage{\ATLASLATEXPATH atlasbsm} 

\usepackage{\ATLASLATEXPATH atlasphysics}

\graphicspath{{logos/}{figures/}}

\usepackage{HIGG-2017-04-PAPER-defs}

\input{HIGG-2017-04-PAPER-metadata}

\hypersetup{pdftitle={ATLAS document},pdfauthor={The ATLAS Collaboration}}

\begin{document}

\maketitle

\tableofcontents

\input{introduction}

\input{detector}

\input{simulationsamples}

\input{objectselection}

\input{analysisstrategy}

\input{systematics}

\input{statistical}

\input{results}

\input{conclusions}

\section*{Acknowledgements}
\input{acknowledgements/Acknowledgements}

\clearpage
\appendix
\part*{Appendix}
\addcontentsline{toc}{part}{Appendix}

\input{appendix}

\clearpage

\printbibliography

\clearpage \input{atlas_authlist}

\end{document}

%% file: HIGG-2017-04-PAPER-metadata.tex
\AtlasTitle{Search for charged Higgs bosons decaying into top and bottom quarks at $\sqrt{s}=13$~\TeV\ with the ATLAS detector}

\AtlasAbstract{%
A search for charged Higgs bosons heavier than the top quark and decaying via $H^\pm\to tb$ is presented. 
The data analysed corresponds to 36.1 fb$^{-1}$ of $pp$ collisions at $\sqrt{s}=13$ \TeV\ and was recorded with the ATLAS detector at the LHC in 2015 and 2016. 
The production of a charged Higgs boson in association with a top quark and a bottom quark, $pp \to tbH^\pm$, is explored in the mass range from $m_{H^\pm}=200$ to $2000$~\GeV\ using multi-jet final states with one or two electrons or muons. 
Events are categorised according to the multiplicity of jets and how likely these are to have originated from hadronisation of a bottom quark. 
Multivariate techniques are used to discriminate between signal and background events. 
No significant excess above the background-only hypothesis is observed and exclusion limits are derived for the production cross-section times branching ratio of a charged Higgs boson as a function of its mass, which range from 2.9~pb at $m_{H^\pm} = 200$~\GeV\ to 0.070~pb at $m_{H^\pm} = 2000$~\GeV. 
The results are interpreted in two benchmark scenarios of the Minimal Supersymmetric Standard Model.
}

\author{The ATLAS Collaboration}

\AtlasRefCode{HIGG-2017-04}

\PreprintIdNumber{CERN-EP-2018-168}

\AtlasJournalRef{JHEP 11 (2018) 085} 

\AtlasDOI{10.1007/JHEP11(2018)085}

 \arXivId{1808.03599}

\HepDataRecord{https://www.hepdata.net/record/83203}

\AtlasJournal{JHEP}

%% file: introduction.tex
\section{Introduction}
\label{sec:intro}

Following the discovery of a Higgs boson, \Higgs, with a mass of around 125~\GeV\
and consistent with the Standard Model (SM)~\cite{HIGG-2012-27,Chatrchyan:2012ufa,HIGG-2014-14}
at the Large Hadron Collider (LHC) in 2012~\cite{Evans:2008zzb}
a key question is whether this Higgs boson is the only Higgs boson, or the first observed physical state of an extended Higgs sector.
No charged fundamental scalar boson exists in the SM, but many beyond the Standard Model (BSM) scenarios
contain an extended Higgs sector with at least one set of charged Higgs bosons, \hplus\ and \hminus, 
in particular two-Higgs-doublet models (2HDM)~\cite{Lee:1973iz,Akeroyd:2016ymd,Gunion:2002zf,Branco:2011iw} 
and models containing Higgs triplets~\cite{trip1,trip2,trip3,trip4,trip5}.

\begin{figure}[htbp]
\begin{center}
\includegraphics[width=0.3\textwidth]{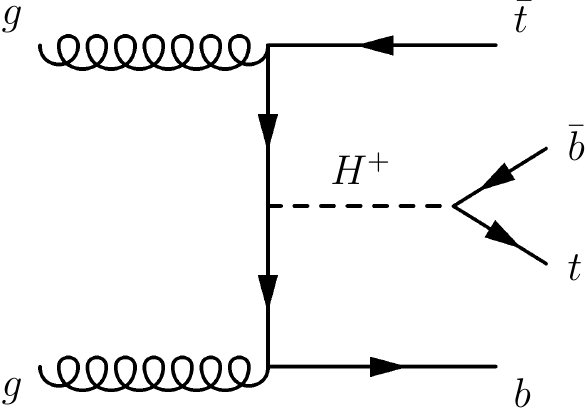}
\end{center}
\caption{Leading-order Feynman diagram for the production of a heavy charged Higgs boson ($\mhp > m_t + m_b$)
in association with a top quark and a bottom quark (\tbhplus ). 
\label{fig:hplus_feynman_production}}
\end{figure}

The production mechanisms and decay modes of a charged Higgs boson\footnote{For simplicity in the following, charged Higgs bosons are denoted \hplus, with the charge-conjugate \hminus\ always implied. 
Similarly, the difference between quarks and antiquarks, $q$ and $\bar{q}$, is generally understood from the context, 
so that e.g. $\hplus\to tb$ means both $\hplus\to t\bar{b}$ and $\hminus\to\bar{t}b$.}
 depend on its mass, \mhp.
This analysis searches for heavy charged Higgs bosons with $\mhp > m_t + m_b$, where $m_t$ and $m_b$ are the masses of the top and bottom quarks, respectively.
The dominant production mode is expected to be in association with a top quark and a bottom quark (\tbhplus),
as illustrated in Figure~\ref{fig:hplus_feynman_production}. 
In the 2HDM, \hplus\ production and decay at tree level depend on its mass and two parameters:
\tanbeta\ and $\alpha$, which are the ratio of the vacuum expectation values of the two Higgs doublets
and the mixing angle between the CP-even Higgs bosons, respectively.
The dominant decay mode for heavy charged Higgs bosons is $\hplus \to tb$ in a broad range of models~\cite{Heinemeyer:2013tqa,deFlorian:2016spz}. In particular, this is the preferred decay mode in both the decoupling limit scenario and the alignment limit  $\cos(\beta - \alpha)\approx 0$, where the lightest CP-even neutral Higgs boson of the extended Higgs sector has properties similar to those of the SM Higgs boson~\cite{Gunion:2002zf}.
For lower \mhp, the dominant decay mode is $\hplus \to \tau \nu$.
It is also predicted that this decay mode becomes more relevant as the value of \tanbeta\ increases, irrespective of \mhp.
Therefore, the $\hplus \to tb$ and $\hplus \to \tau \nu$ decays naturally complement each other in searches for charged Higgs bosons.

Limits on charged Higgs boson production have been obtained by many experiments,
such as the LEP experiments 
with upper limits on \hplus\ production in the mass range 40--100~\GeV~\cite{Abbiendi:2013hk}, 
and CDF and D\O{} at the Tevatron 
that set upper limits on the branching ratio $\BR(t \to b\hplus)$ for $80\, \GeV < \mhp < 150\, \GeV$~\cite{Aaltonen:2009ke,Abazov:2009aa}. 
The CMS Collaboration has performed direct searches for heavy charged Higgs bosons in 8~\TeV\ proton--proton ($pp$) collisions. By assuming the branching ratio $\BR(\hplus \to tb) = 1$, an upper limit of 2.0--0.13~pb was obtained for the production cross-section $\sigma(pp \to tb \hplus)$ for $180\, \GeV \, < \mhp < 600\, \GeV$~\cite{Khachatryan:2015qxa}.  
The ATLAS Collaboration has searched for similar heavy charged Higgs boson production in the $\hplus \to tb$ decay channel at 8~\TeV, setting upper limits on the production cross-section times the $\hplus \to tb$ branching ratio of 6--0.2~pb for $200\, \GeV \, < \mhp < 600\, \GeV$~\cite{Aad:2015typ}.
Indirect constraints can be obtained from the measurement of flavour-physics observables sensitive to charged Higgs boson exchange. Such observables include the relative branching ratios of $B$ or $K$ meson decays, $B$ meson mixing parameters, the ratio of the $Z$ decay partial widths $\Gamma( Z \to b\bar{b} ) / \Gamma( Z \to \text{hadrons} )$, as well as the measurements of $b \to s\gamma$ decays~\cite{Deschamps:2009rh, Arbey:2017gmh}. The relative branching ratio $\mathcal{R}(D^{(*)}) = \BR( B \to D^{(*)}\tau\nu ) / \BR( B \to D^{(*)}\ell\nu )$, where $\ell$ denotes $e$ or $\mu$, are especially sensitive to contributions from new physics. 
Measurements from BaBar~\cite{Lees:2013uzd} exclude \hplus\ for all \mhp\ and \tanbeta\ values in a Type-II 2HDM. 
However, more recent measurements from Belle~\cite{Hirose:2017dxl,Sato:2016svk,Huschle:2015rga} and LHCb~\cite{Aaij:2017deq} place a weaker constraint on the allowed range of $\mhp/\tanbeta$ values. 
A global fit combining the most recent flavour-physics results~\cite{Arbey:2017gmh} sets a lower limit at 95\% confidence level on the charged Higgs boson mass of $\mhp \gtrsim 600\,\GeV$  for $\tanbeta > 1$ and $\mhp \gtrsim 650\,\GeV$ for lower \tanbeta\ values, assuming a Type-II 2HDM. 

This paper presents a search for \hplus\ production in the \hplustb\ decay mode using $pp$ collisions at $\sqrt{s}=13$~\TeV.
Events with one charged lepton ($\ell = e, \mu$) and jets in the final state (\ljets\ final state) 
and events with two charged leptons and jets in the final state (\dilepton\ final state) are considered. 
Exclusive regions are defined according to the number of jets and those that are tagged as originating from the hadronisation of a $b$-quark. 
In order to separate the signal from the SM background, multivariate discriminants are employed in the regions where the   signal contributions are expected to be largest. 
Limits on the \hplustb\ production cross-section are set by means of a simultaneous fit of binned distributions of multivariate discriminants in the signal-rich regions and inclusive event yields in the signal-depleted regions.
The results are interpreted in two benchmark scenarios of the Minimal Supersymmetric Standard Model (MSSM): the \mhmodm\ scenario~\cite{Carena:2013ytb} and the hMSSM~\cite{Djouadi:2013uqa}. 
Both scenarios exploit the MSSM in such a way that the light CP-even Higgs boson can be interpreted as the observed Higgs boson with $m_\Higgs = 125$~\GeV.
Limits on the value of \tanbeta\ are extracted as a function of the charged Higgs boson mass.
Finally, the excluded range of \mhp\ and \tanbeta\ values from the $\hplus \to tb$ and $\hplus \to \tau \nu$~\cite{ATLAS-HIGG-2016-11} searches at $\sqrt{s}=13$~\TeV\ are superimposed,
providing a summary of the ATLAS sensitivity to \hplus\ through the two decay modes.

The paper is organised as follows.
Section~\ref{sec:detector} briefly describes the ATLAS detector. 
The samples of simulated events used for the analysis are summarised in Section~\ref{sec:samples}. 
Section~\ref{sec:selection} presents the reconstruction of objects in ATLAS and the event selection.
Section~\ref{sec:analysis} describes the analysis strategy while systematic uncertainties are discussed in Section~\ref{sec:systematics}.
The statistical analysis of the data is described in Section~\ref{sec:statistical} and the results are presented in
Section~\ref{sec:result}. 
Finally, a summary is given in Section~\ref{sec:conclusion}.

%% file: detector.tex
\section{ATLAS detector}
\label{sec:detector}

The ATLAS detector~\cite{PERF-2007-01} at the LHC is a multipurpose particle detector with a forward--backward symmetric cylindrical geometry and near $4\pi$ coverage around the collision point.\footnote{
ATLAS uses a right-handed coordinate system with its origin at the nominal interaction point (IP) in the centre of the detector and the $z$-axis along the beam pipe. 
The $x$-axis points from the IP to the centre of the LHC ring, and the $y$-axis points upwards. 
Cylindrical coordinates $(r,\phi)$ are used in the transverse plane, $\phi$ being the azimuthal angle around the $z$-axis. 
The pseudorapidity is defined in terms of the polar angle $\theta$ as $\eta = -\ln \tan(\theta/2)$. 
Angular distance is measured in units of $\Delta R \equiv \sqrt{(\Delta\eta)^{2} + (\Delta\phi)^{2}}$ (pseudorapidity and azimuthal angle). 
Alternatively, the distance $\Delta R_y \equiv \sqrt{(\Delta y)^{2} + (\Delta\phi)^{2}}$ is used, where 
$y= 0.5 \ln \left[ \left(E+p_z\right) / \left(E-p_z\right) \right]$ 
is the rapidity of a particle of energy $E$ and momentum component $p_z$ along the beam axis.
} 
The ATLAS detector consists of an inner tracking detector (ID) surrounded by a thin superconducting solenoid producing a 2~T axial magnetic field, electromagnetic (EM) and hadronic calorimeters, and an external muon spectrometer (MS) incorporating three large toroid magnet assemblies. 
The ID contains a high-granularity silicon pixel detector, including an insertable B-layer~\cite{IBL} added in 2014 as a new innermost layer,
and a silicon microstrip tracker, providing precision tracking in the pseudorapidity range $|\eta|<2.5$. The silicon detectors are complemented by a transition radiation tracker providing tracking and electron identification information for $|\eta|<2.0$. 
The EM sampling calorimeter uses lead as the absorber material and liquid argon (LAr) as the active medium, and is divided into barrel ($|\eta|<1.47$) and endcap ($1.37<|\eta|<3.20$) regions. Hadron calorimetry is also based on the sampling technique, with scintillator tiles or LAr as the active medium, and with steel, copper, or tungsten as the absorber material. The calorimeters cover $|\eta|<4.9$. The MS measures the deflection of muons with $|\eta|<2.7$ using multiple layers of high-precision tracking chambers located in a toroidal field in the central and endcap regions of ATLAS.
The field integral of the toroids ranges between \num{2.0} and \SI{6.0}{\tesla\metre} across most of the detector.
The MS is also instrumented with separate trigger chambers covering $|\eta|<2.4$.
A two-level trigger system, with the first level implemented in custom hardware and followed by a software-based second level, is used to reduce the trigger rate to around 1~kHz for offline storage~\cite{trigger2015}.

%% file: simulationsamples.tex
\section{Signal and background modelling}
\label{sec:samples}

The \tbhplus\ process was modelled with \MGMCatNLO\ (\mg5\_\amc)~\cite{Alwall:2014hca} at next-to-leading order (NLO) in QCD~\cite{Degrande:2015vpa} using a four-flavour scheme (4FS) implementation with the \nnpdf 2.3NLO~\cite{Ball:2012cx}\ parton distribution function (PDF).\footnote{Five-flavour scheme (5FS) PDFs consider $b$-quarks as a source of incoming partons and the $b$-quarks are therefore assumed to be massless. In contrast, 4FS PDFs only include lighter quarks and gluons, allowing the $b$-quark mass to be taken into account properly in the matrix element calculation.}
Parton showering and hadronisation were modelled by \PYTHIAV{8.186}~\cite{Sjostrand:2007gs} 
with the \textsc{A14}~\cite{a14} set of underlying-event (UE) related parameters tuned to ATLAS data (tune). 
For the simulation of the \tbhplus\ process, the narrow-width approximation was used. 
This assumption has a negligible impact on the analysis for the models considered in this paper, as the experimental resolution is much larger than the \hplus\ natural width.
Interference with the SM $\ttbar+\bbbar$ background is neglected. 

Altogether 18 \hplus\ mass hypotheses are used, with 25~\GeV\ mass steps between an \hplus\ mass of 200~\GeV\ and 300~\GeV,
50~\GeV\ steps between 300~\GeV\ and 400~\GeV,
100~\GeV\ steps between 400~\GeV\ and 1000~\GeV\ and 200~\GeV\ steps from 1000~\GeV\ to 2000~\GeV. 
The step sizes are selected to match the expected resolution of the \hplus\ signal. 
The samples were processed with a 
fast simulation of the ATLAS detector~\cite{fastsim}. 
Unless otherwise indicated, the cross-section of the signal is set to 1~pb, for easy rescaling to various model predictions. 
Only the \hplus\ decay into $tb$ is considered, and the top quark decays according to the SM predictions. 

The nominal sample used to model the \ttbar\ background was generated using the \POWHEGBOXV{v2} NLO-in-QCD generator~\cite{Nason:2004rx,Frixione:2007vw,Alioli:2010xd,Campbell:2014kua},
referred to as \powheg\ in the remainder of this article,
with the \nnpdf3.0NLO PDF set~\cite{Ball:2014uwa}. 
The \hdamp\ parameter, which controls the transverse momentum \pt\ of the first additional emission beyond the Born configuration, was set to 1.5 times the top quark mass~\cite{ATL-PHYS-PUB-2016-020}. 
Parton shower and hadronisation were modelled by \PYTHIAV{8.210}~\cite{Sjostrand:2014zea} with the \textsc{A14} UE tune. 
The sample was normalised to the \toppp2.0~\cite{Czakon:2011xx} theoretical cross-section of $832^{+46}_{-51}$~pb, calculated at next-to-next-to-leading order (NNLO) in QCD 
including resummation of next-to-next-to-leading logarithmic (NNLL) soft gluon terms~\cite{ref:xs1,ref:xs2,ref:xs3,ref:xs4,ref:xs5}. 
The generation of the \ttbar\ sample was performed inclusively, with all possible flavours of additional jets produced. 
The decay of $c$- and $b$-hadrons was simulated with the \evtgen\ v1.2.0~\cite{Lange:2001uf} program. 
The \ttbar + jets background is categorised according to the flavour of additional jets in the event, using the same procedure as described in Ref.~\cite{HIGG-2013-27}. 
The \ttbar+ additional heavy-flavour (HF) jets background is subdivided into the categories \ttgeb\ and \ttgec, depending on whether the additional HF jets originate from hadrons containing $b$- or $c$-quarks. 
Particle jets were reconstructed from stable particles (mean lifetime $\tau > 3\times 10^{-11}$~seconds) at generator level using the anti-\kt\ algorithm~\cite{Cacciari:2008gp} with a radius parameter of 0.4, and were required to have $\pt > 15\,\GeV$ and $|\eta| < 2.5$. 
If at least one particle-level jet in the event is matched ($\Delta R<0.3$) to a $b$-hadron (not originating from a $t$-decay) with $\pt>5$~\GeV, the event is categorised as \ttgeb. 
In the remaining events, if at least one jet is matched to a $c$-hadron (not originating from a $W$~decay) but no $b$-hadron, the event is categorised as \ttgec. 
Events with \ttbar + jets that belong to neither the \ttgeb\ nor \ttgec\ category are called \ttbar + light events. 

For the \ttbin\ process, subcategories are defined in accord with
the matching between particle-level jets and the $b$-hadrons not from $t$-decay:
events where exactly two jets are matched to $b$-hadrons (\ttbb), events where exactly one jet is matched to a $b$-hadron (\ttb), events where exactly one jet is matched to two or more $b$-hadrons (\ttB), and all other events (\ttbbb).
Events where the additional HF jets can only be matched to $b$-hadrons from multi-parton interactions 
and final-state gluon radiation are considered separately and labelled as \ttb\ (MPI/FSR).

To model the irreducible \ttbin\ background to the highest available precision, the \ttbin\ events from the nominal \powheg+\pythia8 simulation are reweighted to an NLO prediction of $\ttbar\bbbar$ including parton showering and hadronisation from \SHERPAV{2.1.1}~\cite{Gleisberg:2008ta,Cascioli:2013era} with \ol~\cite{Cascioli:2011va}.
This sample was generated using the 4FS PDF set CT10F4~\cite{Lai:2010vv}.
The renormalisation scale (\muR) for this sample was set to the $\mu_\mathrm{CMMPS} = \prod_{i=t,\bar{t},b,\bar{b}} E_{\mathrm{T},i}^{1/4}$~\cite{Cascioli:2013era,MorettiPhD},
and the factorisation (\muF) and resummation (\muQ) scales to $H_\mathrm{T}/2=\frac{1}{2}\sum_{i=t,\bar{t},b,\bar{b}} E_{\mathrm{T},i}$.
A first type of reweighting is performed in the \ttgeb\ subcategories, using a method similar to the one outlined in Ref.~\cite{Aaboud:2017rss}.
The reweighting corrects the relative normalisation of the \ttgeb\ subcategories to match the predictions from \SHERPA, while keeping the overall \ttgeb\ normalisation unchanged. 
After applying the first reweighting based on the relative normalisation of the \ttgeb\ subcategories, a second type of reweighting is derived and performed on several kinematic variables sequentially.
First the \pt\ of the \ttbar\ system is reweighted, and secondly the \pt\ of the top quarks. 
The final reweighting is performed depending on the type of \ttgeb\ events. 
If there is only one additional HF jet, the \pt\ of that jet is used in the final reweighting. 
If there is more than one additional HF jet, first the $\Delta R$ between the HF jets is reweighted and then the \pt\ of the HF dijet system.
A closure test is performed on each of the reweighted kinematic variables,
showing a reasonable level of agreement between the reweighted \powheg+\pythia8 sample and the \SHERPA sample.

The \POWHEGBOXV{v1} generator was used to produce the samples of $Wt$ single-top-quark backgrounds, with the \textsc{CT10} PDF set. 
Overlaps between the \ttbar\ and $Wt$ final states were handled using the `diagram removal' scheme~\cite{Frixione:2008yi}. 
The $t$-channel single-top-quark events were generated using the \POWHEGBOXV{v1} generator with the 4FS for the NLO matrix element calculations and the fixed 4FS PDF set CT10F4. 
The top quarks were decayed with \madspin~\cite{Artoisenet:2012st}, which preserves the spin correlations. 
The samples were interfaced to \PYTHIAV{6.428}~\cite{Sjostrand:2006za} with the \perugia\ 2012 UE tune~\cite{Skands:2010ak}. 
The single-top-quark $Wt$ and $t$-channel samples were normalised to the approximate NNLO (aNNLO) theoretical cross-section~\cite{Kidonakis:2010ux,Kidonakis:2010tc,Kidonakis:2011wy}.

Samples of $W/Z$+jets events were generated using \SHERPAV{2.2.1}~\cite{Gleisberg:2008ta}. 
Matrix elements were calculated for up to 2 partons at NLO and 4 partons at LO using \comix~\cite{Gleisberg:2008fv} 
and \ol\ and merged with the \SHERPA parton shower~\cite{Schumann:2007mg} using the ME+PS@NLO prescription~\cite{Hoeche:2012yf}. 
The NNPDF3.0NNLO PDF set was used together with a dedicated parton shower tune developed by the \SHERPA authors. 
The $W/Z$+jets events were normalised to the NNLO cross-sections~\cite{Melnikov:2006kv,Gavin:2010az,Li:2012wna,Bardin:2012jk,Arbuzov:2012dx}.

Samples of \ttV\ ($V= W,\,Z$) events were generated at NLO in the matrix elements calculation using \mg5\_\amc\ with the \nnpdf3.0NLO PDF set interfaced to \PYTHIAV{8.210} with the A14 UE tune. 
The \ttH\ process was modelled using \mg5\_\amc\ with NLO matrix elements, \nnpdf3.0NLO PDF set and factorisation and renormalisation scales set to $\muF = \muR = \mtrans / 2$, where \mtrans\ is defined as the scalar sum of the transverse masses $\mtrans = \sqrt{\pT^2 + m^2}$ of all final-state particles. 
The events were interfaced to \PYTHIAV{8.210} with the A14 UE tune.
Variations in \ttH\ production due to the extended Higgs sector are not considered in this analysis, since the contribution from the \ttH\ background is found to be small.
Measurements of the \ttH\ production cross-section are compatible with the SM expectation~\cite{Aaboud:2018urx,Sirunyan:2018hoz}.

The minor $\tH +X$ backgrounds, consisting of the production of a single top quark in association with a Higgs boson and jets (\tHjb), 
and the production of a single top quark, a $W$ boson and a Higgs boson (\WtH), are treated as one background.
The \tHjb\ process was simulated with \mg5\_\amc\ interfaced to \PYTHIAV{8.210} and the CT10 PDF set, 
and \WtH\ was modelled with \mg5\_\amc\ interfaced to \HERWIGpp~\cite{Bahr:2008pv} using the CTEQ6L1 PDF set~\cite{cteq6l1}. 
Additional minor SM backgrounds (diboson production, single top $s$-channel, $tZ$, $tWZ$, $4t$, $ttWW$) were also simulated and accounted for, even though they contribute less than 1\% in any analysis region. 

Except where otherwise stated, all simulated event samples were produced using the full ATLAS detector simulation~\cite{Aad:2010ah} based on \GEANT 4~\cite{Agostinelli:2002hh}. 
Additional pile-up interactions were simulated with \PYTHIAV{8.186} using the A2 set of tuned parameters~\cite{ATLAS:2012uec} and the MSTW2008LO PDF set~\cite{Martin:2009iq},
and overlaid onto the simulated hard-scatter event. 
All simulated samples were reweighted such that the average number of interactions per bunch crossing (pile-up) matches that of the data. 
In the simulation, the top quark mass was set to $m_t = 172.5\,\GeV$. Decays of $b$- and $c$-hadrons were performed by \evtgen\ v1.2.0, except in samples simulated by the \SHERPA event generator.

The samples and their basic generation parameters are summarised in Table~\ref{tab:simulation_summary}. 

\begin{table}[h!]
\caption{Nominal simulated signal and background event samples. The generator, parton shower generator and cross-section used for normalisation are shown together with the applied PDF set and tune. The $\ttbar\bbbar$ event sample generated using \SHERPAV{2.1.1} is used to reweight the events from the \ttbin\ process in the \ttbar + jets sample.\label{tab:simulation_summary}}
  \footnotesize
\begin{center}
    \begin{tabular}{l l l l l l}

Physics process                  & Generator       & Parton shower   & Cross-section   & PDF set       & Tune \\
                                 &                 & generator       & normalisation   &               &      \\
\hline\hline
 \tbhplus                        & \mg5\_\amc      & \PYTHIAV{8.186} & --              & \nnpdf2.3NLO  & \textsc{A14}\\
\hline
\ttbar + jets                    & \POWHEGBOXV{v2} & \PYTHIAV{8.210} & NNLO+NNLL       & \nnpdf3.0NLO  & \textsc{A14}\\
$\ttbar\bbbar$                   & \SHERPAV{2.1.1} & \SHERPAV{2.1.1} & NLO for $\ttbar\bbbar$ & CT10F4 & \SHERPA default \\
 \ttV                            & \mg5\_\amc      & \PYTHIAV{8.210} & NLO             & \nnpdf3.0     & \textsc{A14}\\
 \ttH                            & \mg5\_\amc      & \PYTHIAV{8.210} & NLO             & \nnpdf3.0NLO  & \textsc{A14}\\
Single top, $Wt$ 
                                 & \POWHEGBOXV{v1} & \PYTHIAV{6.428} & aNNLO           & CT10          & Perugia 2012 \\  
Single top, $t$-channel          & \POWHEGBOXV{v1} & \PYTHIAV{6.428} & aNNLO           & CT10F4        & Perugia 2012 \\
\hline
 $W$+jets                        & \SHERPAV{2.2.1} & \SHERPAV{2.2.1} & NNLO            & \nnpdf3.0NNLO & \SHERPA default \\
 $Z$+jets                        & \SHERPAV{2.2.1} & \SHERPAV{2.2.1} & NNLO            & \nnpdf3.0NNLO & \SHERPA default \\
    \end{tabular}
\end{center}
\end{table}

%% file: objectselection.tex
\section{Object and event selection}
\label{sec:selection}

The data used in this analysis were recorded in 2015 and 2016 from $\sqrt{s}=13$~\TeV\ $pp$ collisions with an integrated luminosity of \lumi. 
Only runs with stable colliding beams and in which all relevant detector components were functional are used.
Events are required to have at least one reconstructed vertex with two or more tracks with $\pt > 0.4$~\GeV.
The vertex with the largest sum of the squared \pt\ of associated tracks is taken as the primary vertex.

Events were recorded using single-lepton triggers, in both the \ljets\ and \dilepton\ final states. 
To maximise the event selection efficiency, multiple triggers were used, with either low \pt\ thresholds and lepton identification and isolation requirements, or with higher \pt\ thresholds but looser identification criteria and no isolation requirements. 
Slightly different sets of triggers were used for 2015 and 2016 data. 
For muons, the lowest \pt\ threshold was 20 (26)~\GeV\ in 2015 (2016), 
while for electrons, triggers with a \pt\ threshold of 24 (26)~\GeV\ were used. 
Simulated events were also required to satisfy the trigger criteria.

Electrons are reconstructed from energy clusters in the EM calorimeter associated with tracks reconstructed in the ID~\cite{ATLAS-CONF-2016-024}.
Candidates in the calorimeter transition region $1.37 <|\eta_{\mathrm{cluster}}| < 1.52$ are excluded.
Electrons are required to satisfy the tight identification criterion described in Ref.~\cite{ATLAS-CONF-2016-024}, based on shower-shape and track-matching variables. 
Muons are reconstructed from track segments in the MS that are matched to tracks in the ID~\cite{Aad:2016jkr}.
Tracks are then re-fit using information from both detector systems.
The medium identification criterion described in Ref.~\cite{Aad:2016jkr} is used to select muons.
To reduce the contribution of leptons from hadronic decays (non-prompt leptons), 
both the electrons and muons must satisfy isolation criteria. 
These criteria include both track and calorimeter information, 
and have an efficiency of 90\% for leptons with a \pt\ of 25~\GeV, rising to 99\% above 60~\GeV, as measured in $Z \to ee$~\cite{ATLAS-CONF-2016-024} and $Z \to \mu\mu$~\cite{Aad:2016jkr} samples. 
Finally, the lepton tracks must point to the primary vertex of the event:
the longitudinal impact parameter $z_0$ must satisfy $|z_0 \mathrm{sin}\theta| < 0.5$~mm,
while the transverse impact parameter significance must satisfy, 
$|d_{0}|/\sigma(|d_{0}|) < 5$ (3) for electrons (muons).

Jets are reconstructed from three-dimensional topological energy clusters~\cite{PERF-2014-07}
 in the calorimeter using the anti-\kt\ jet algorithm~\cite{Cacciari:2008gp,Fastjet} with a radius parameter of 0.4. 
Each topological cluster is calibrated to the EM scale response prior to jet reconstruction.
The reconstructed jets are then calibrated to the jet energy scale (JES) derived from simulation and \emph{in situ} corrections based on $\sqrt{s}=13$~\TeV\ data~\cite{PERF-2016-04}. 
After energy calibration, jets are required to have $\pt > 25\,\GeV$ and $|\eta| < 2.5$. 
Quality criteria are imposed to identify jets arising from non-collision sources or detector noise, 
and events containing any such jets are removed~\cite{Run2-jet-cleaning}. 
Finally, to reduce the effect of pile-up an additional requirement using information about the tracks and the primary vertex associated to a jet (Jet Vertex Tagger)~\cite{Aad:2015ina} is applied for jets with $\pt < 60$~\GeV\ and $|\eta| < 2.4$. 

Jets are identified as containing the decay of a $b$-hadron ($b$-tagged) via an algorithm using multivariate techniques to combine information from the impact parameters of displaced tracks with the topological properties of secondary and tertiary decay vertices reconstructed within the jet~\cite{Aad:2015ydr,ATL-PHYS-PUB-2016-012}. 
Jets are $b$-tagged by directly requiring the output discriminant of the $b$-tagging algorithm to be above a threshold. 
A criterion with an efficiency of 70\% for $b$-jets in \ttbar\ events is used to determine the $b$-jet multiplicity for all final states and \hplus\ masses. 
For this working point, the $c$-jet and light-jet rejection factors are 12 and 381, respectively. 
For $\mhp \leq 300$~\GeV, 
five exclusive efficiency bins are defined using the same $b$-tagging discriminant: 0--60\%, 60--70\%, 70--77\%, 77--85\% and 85--100\%, following the procedure described in Ref.~\cite{HIGG-2013-23}.
These step-wise efficiencies are used as input to the kinematic discriminant described in Section~\ref{sec:analysis}.
When `a $b$-tagged jet' is mentioned without any further specification, an efficiency of 70\% is implied. 

To avoid counting a single detector response as two objects, an overlap removal procedure is used.
First, the closest jet within $\Delta R_y = 0.2$ of a selected electron is removed.
If the nearest jet surviving this selection is within
$\Delta R_y = 0.4$ of the electron, the electron is discarded,
to ensure it is sufficiently separated from nearby jet activity.
Muons are removed if they are separated from the nearest jet by $\Delta R_y < 0.4$,
to reduce the background from muons from HF decays inside jets.
However, if this jet has fewer than three associated tracks,
the muon is kept and the jet is removed instead;
this avoids an inefficiency for high-energy muons undergoing significant energy loss in the calorimeter.

The missing transverse momentum in the event is defined as the negative vector sum of the \pt\ of all the selected electrons, muons and jets described above, with an extra term added to account for energy in the event that is not associated with any of these. 
This extra term, referred to as the `soft term' in the following, is calculated from ID tracks matched to the primary vertex to make it resilient to pile-up contamination~\cite{PERF-2014-04,PERF-2016-07,ATLAS-CONF-2018-023}. 
The missing transverse momentum is not used for event selection but is an input to the multivariate discriminants.

Events are required to have at least one electron or muon. 
The leading lepton must be matched to a lepton with the same flavour reconstructed by the trigger algorithm within $\Delta R < 0.15$, and have a $\pt > 27$~\GeV. 
Additional leptons are required to have $\pt> 10$~\GeV, or $> 15$~\GeV\ for events with two electrons.
The latter requirement reduces the bakground due to jets and photons that are misidentified as electrons.
Events in the \ljets\ channel and the \dilepton\ channel are required to be mutually exclusive. 
Electrons or muons from $\tau$ decays are also included in the analysis. 

For the \ljets\ channel, five or more jets, of which at least two jets have to be $b$-tagged, are required. 
For the \dilepton\ channel, events with two leptons with opposite charge are selected, and at least three jets are required, of which two or more must be $b$-tagged. 
In the $ee$ and $\mu\mu$ channels, the dilepton invariant mass must be $ > 15$~\GeV\ and outside the $Z$ boson mass window of 83--99 \GeV.

%% file: analysisstrategy.tex
\section{Analysis strategy}
\label{sec:analysis}
\
After the event selection, the samples in both the \dilepton\ and the \ljets\ final states contain mostly \ttbar\ events. 
Events passing the event selection are categorised into separate regions according to the number of reconstructed jets and $b$-tagged jets. 
The regions where $tb\hplus$ is enhanced relative to the backgrounds are referred to as signal regions (SRs), whereas the remaining regions are referred to as control regions (CRs). 

For the \ljets\ final state, two CRs (\FiveJexTwoBex\ and \SixJinTwoBex)\footnote{$XjYb$ means that $X$ jets are found in the event, 
and among them $Y$ are $b$-tagged.} and four SRs (\FiveJexThreeBex, \FiveJexFourBin, \SixJinThreeBex\ and \SixJinFourBin) are defined, while in the \dilepton\ final state, two CRs (\ThreeJexTwoBex\ and \FourJinTwoBex) 
and two SRs (\FourJinThreeBex\ and \FourJinFourBin) are defined for all mass hypotheses. 
In addition, for the  \dilepton\ final state, the region with three $b$-tagged jets and no other jets (\ThreeJexThreeBex) is considered a SR for $\mhp<1\,\TeV$ and a CR for $\mhp \geq 1\,\TeV$
due to the change in expected signal yield for the different \hplus\ mass hypotheses.
 
In the SRs, for each \hplus\ mass hypothesis a different discriminating variable based on boosted decision trees (BDTs) is defined. 
In order to separate the \hplus\ signal from the SM background, the binned output of this variable is used together with the total event yields in the CRs in a combined profile likelihood fit.
The fit simultaneously determines both the signal and background yields, while constraining the overall background model within the assigned systematic uncertainties. 
The event yields in the CRs are used to constrain the background normalisation and systematic uncertainties.
In the following subsections the background estimate and the design of the multivariate discriminator are described. The profile likelihood fit, including the treatment of backgrounds in the fit, is described in detail in Section~\ref{sec:statistical}.

\subsection{Background estimate}
\label{backgroundestimate}
The background from processes with prompt leptons is estimated using the simulated event samples described in Section~\ref{sec:samples}. 
For \ttbar\ production, the number of events with high leading jet \pt\ is overestimated in the simulation, 
and a reweighting function for the leading jet \pt\ distribution is determined by comparing simulation with data in a \ljets\ CR that requires exactly four jets and at least two $b$-tagged jets. This function is validated in the dilepton channel and applied to both channels.

The normalisation of the $Z$+HF jets backgrounds is corrected by a factor of $1.3$, extracted from dedicated control regions in data, defined by requiring two opposite-charge same-flavour leptons ($e^+e^-$ or $\mu^+\mu^-$) with an invariant mass compatible with the $Z$ boson mass, $83\ \GeV < m_{\ell\ell} < 99\ \GeV$. 

Processes that do not contain enough prompt electrons or muons from $W$ or $Z$ boson decays can still satisfy the selection criteria if they contain non-prompt leptons. 
The leading sources of non-prompt leptons in the \ljets\ final state are from semileptonic hadron decays or misidentified jets in multi-jet production. In the \dilepton\ final state, the dominant source of non-prompt leptons is from misidentified jets as leptons arising from $W$+jets or \ljets\ $t\bar{t}$ production. These backgrounds are estimated using data.
For the \ljets\ final state a matrix method~\cite{TopFakes8TeV} is employed. An event sample that is enriched in non-prompt leptons is selected by using looser isolation or identification requirements for the lepton. These events are then weighted according to the efficiencies for both the prompt and non-prompt leptons to pass the tighter default selection. These efficiencies are measured using data in dedicated CRs. 
In the \dilepton\ final state, this background is estimated from simulations, and the normalisation is determined by comparing data and simulations in a CR of same-sign dilepton events. 
The contribution of multi-jet events to the \dilepton\ final state is found to be negligible.

\begin{figure}[htbp]
\begin{center}
  \includegraphics[width=0.98\textwidth]{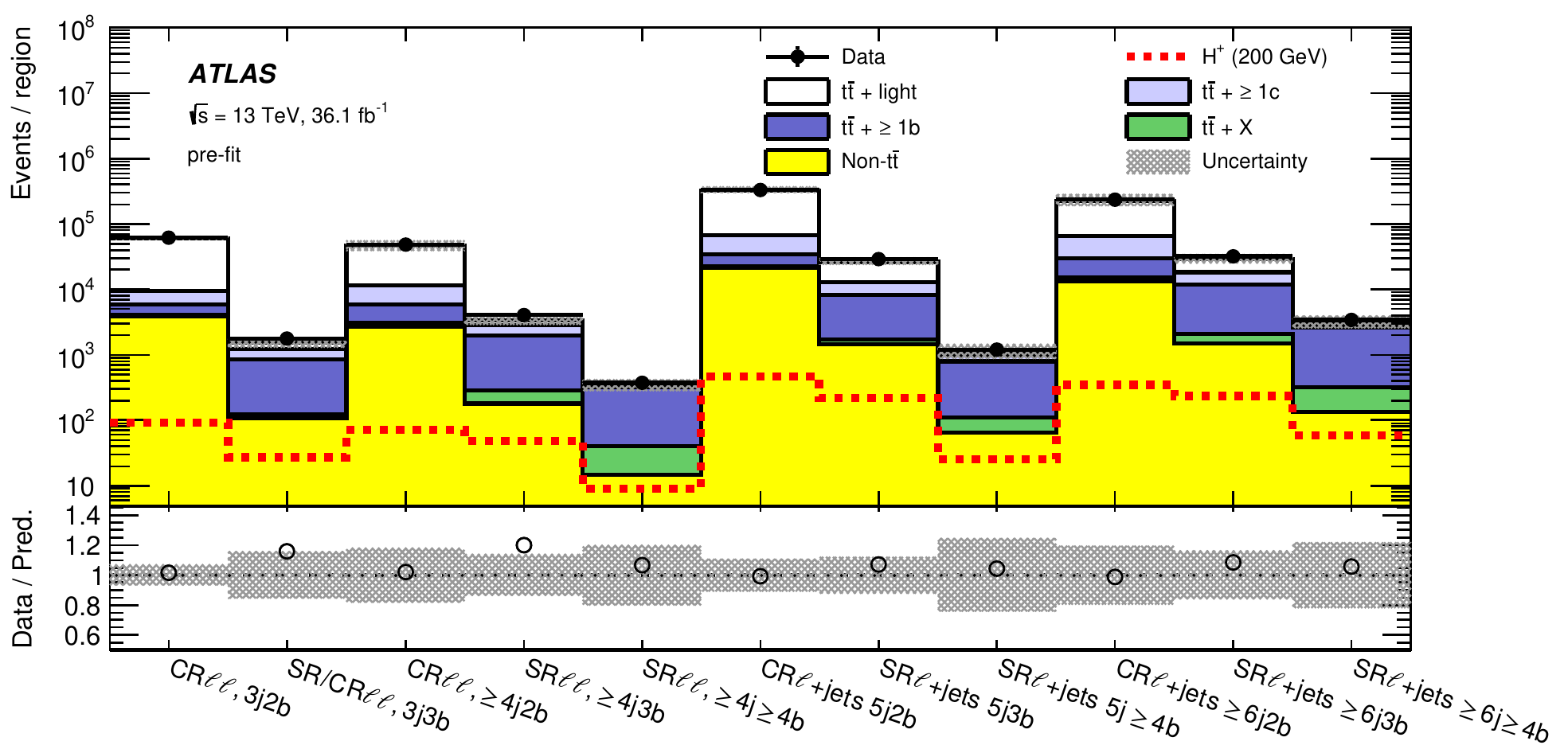}
\end{center}
\caption{
Comparison of predicted and observed event yields.
Each background process is normalised according to its cross-section and the prediction has not been fitted to the data.
The $\ttbar + X$ includes contributions from \ttW, \ttZ\ and \ttH.
A signal with $\mhp = 200\,\GeV$,
normalised to a cross-section times branching ratio for \hplustb\ of 1~pb,
is shown as a dashed line.
The lower panel displays the ratio of the data to the total prediction. 
The hatched bands show uncertainties before the fit to the data, which are dominated by systematic uncertainties as discussed in Section~\ref{sec:systematics}. 
The comparison is shown for all signal and control regions used in the analysis. 
For the \dilepton\ final state: 
CR \ThreeJexTwoBex, 
CR/SR \ThreeJexThreeBex,
CR \FourJinTwoBex, 
SR \FourJinThreeBex, 
SR \FourJinFourBin. 
For the \ljets\ final state: 
CR \FiveJexTwoBex, 
SR \FiveJexThreeBex, 
SR \FiveJexFourBin, 
CR \SixJinTwoBex, 
SR \SixJinThreeBex, 
SR \SixJinFourBin. 
\label{fig:leptonDataCom}
}
\end{figure}

The expected event yields of all SM processes and the number of events observed in the data are shown in Figure~\ref{fig:leptonDataCom} for the \dilepton\ and the \ljets\ final states before performing the fit to data.
The expected \hplus\ signal yields for $\mhp = 200\,\GeV$, assuming a cross-section times branching ratio of 1~pb, are also shown.

\subsection{Multivariate analysis}
\label{mva}
The training of the BDTs that are used to discriminate signal from background in the SRs is performed with the TMVA toolkit~\cite{tmva}. BDTs are trained separately for each value of the 18 generated \hplus\ masses and for each SR against all the backgrounds (\ljets\ channel) or the \ttbar\ background (\dilepton\ channel). For the BDT training in the \ljets\ channel, the SRs \FiveJexThreeBex\ and \FiveJexFourBin\ are treated as one region, in order to increase the number of simulated events available for training.

The BDT variables include various kinematic quantities with the optimal discrimination against the \ttbin\ background. For \hplus\ masses above 400~\GeV\ the most important variables in the \ljets\ final state are the scalar sum of the \pt\ of all jets, \Htjets, and the leading jet \pt. 
For a mass at or below 300~\GeV, a kinematic discriminant, $D$, as described below, is used as an input variable for the BDT. 
The kinematic discriminant, $D$, and the invariant mass of the pair of jets that are not $b$-tagged and have the smallest $\Delta R$ are the most important variables in the low mass range. The latter variable is not used in the \FiveJexFourBin\ SR,  where it is not well defined.

The kinematic discriminant, $D$, is a variable reflecting the probability that an event is compatible with the $\hplus \to tb$ and the $\ttbar$ hypotheses, and is defined as $D=P_{\hplus}(\mathbf{x})/(P_{\hplus}(\mathbf{x})+P_{t\bar{t}}(\mathbf{x}))$,
where $P_{\hplus}(\mathbf{x})$ and $P_{t\bar{t}}(\mathbf{x})$ are probability density functions for $\mathbf{x}$ under the signal hypothesis and background (\ttbar) hypothesis, respectively. 
Here, the event variable $\mathbf{x}$ indicates the set of  the missing transverse momentum and the four-momenta of reconstructed electrons, muons and jets.

The probability $P_{\hplus}(\mathbf{x})$ is defined as the product of the probability density functions for each of the reconstructed invariant masses in the event:
\begin{itemize}
\item the mass of the semileptonically decaying top quark, $m_{b_\ell\ell\nu}$, 
\item the mass of the hadronically decaying $W$ boson, $m_{q_1q_2}$, 
\item the difference between the masses of the hadronically decaying top quark and the hadronically decaying $W$ boson $m_{b_hq_1q_2}-m_{q_1q_2}$, and
\item the difference between the mass of the charged Higgs boson and the mass of the leptonically or hadronically decaying top quark, $m_{b_{H^+}b_\ell\ell\nu}-m_{b_\ell\ell\nu}$ or $m_{b_{H^+}b_hq_1q_2}-m_{b_hq_1q_2}$, depending on whether the top quark from the charged Higgs boson decays leptonically or hadronically.
\end{itemize}
In this context $q_1$ or $q_2$ refer to the quarks from the $W$ boson decay, $\ell$ and $\nu$ to the lepton and neutrino from the other $W$ boson decay, $b_h$ to the $b$-quark from the hadronic top quark decay, $b_\ell$ to the $b$-quark from the leptonic top quark decay and $b_{H^+}$ to the $b$-quark directly from the \hplus~decay.
The probability $P_{t\bar{t}}(\mathbf{x})$ is constructed from probability density functions obtained from simulated \ttbar\ events. 
For the SRs with five jets, $P_{t\bar{t}}(\mathbf{x})$ is defined using the same invariant masses as above.
The jet that does not originate from a top quark decay is used instead of ${b_{H^+}}$. 
For the SRs with at least six jets the power of the discriminant is improved by using 
the invariant mass of the two highest-\pt\ jets not originating from the hadronisation of $q_1$, $q_2$, $b_h$ or $b_\ell$
instead of $m_{b_{H^+}b_\ell\ell\nu}-m_{b_\ell\ell\nu}$ or $m_{b_{H^+}b_hq_1q_2}-m_{b_hq_1q_2}$.

The functional form of the probability density functions is obtained from simulation
using the reconstructed masses of jets and leptons matched to simulated partons and leptons for $H^+$ and \ttbar.
The neutrino four-momentum is derived with the assumption that the missing transverse momentum is solely due to the neutrino;
the constraint $m_W^2=(p_\ell+p_\nu)^2$ is used to obtain $p_{\nu,z}$.
If two real solutions exist, they are sorted according to the absolute value of their $p_z$, i.e., $|p_{z,v_1}| < |p_{z,v_2}|$.
In approximately 60\% of the cases $p_{z,v_1}$ is closer than $p_{z,v_2}$ to the generator-level neutrino $p_z$.
Two different probability density functions are constructed, one for each solution, 
and the probability is defined as a weighted average of the two probability density functions. 
The weight is taken as the fraction of the corresponding solution being closer to the generated neutrino $p_z$. 
Also, if no real solution exists, the $p_x$ and $p_y$ components are scaled by a common factor until the discriminant of the quadratic equation is exactly zero, yielding only one solution.

When evaluating $P_{\hplus}(\mathbf{x})$ and $P_{t\bar{t}}(\mathbf{x})$ for the calculation of $D$, all possible parton–jet assignments are considered since the partonic origin of the jets is not known.
In order to suppress the impact from parton–jet assignments that are inconsistent with the correct parton flavours,
a weighted average over all parton-jet assignments is used.
The value of $P_{\hplus}(\mathbf{x})$ and $P_{t\bar{t}}(\mathbf{x})$ for each parton–jet assignment
is weighted with a probability based on the $b$-tagging discriminant value of each jet.
The distribution of the step-wise efficiencies of the $b$-tagging algorithm, as described in Section~\ref{sec:selection}, is used as a probability density function,
with the $b$-jet hypothesis for generated $b$-quarks and the light-jet hypothesis for other generated partons. 
Due to the large number of events in which $q_1$ and $q_2$ cannot be matched to different jets, the average of two different probability density functions, where either all partons can be matched to jets or only one jet can be matched to $q_1$ and $q_2$, is used. 
This discriminant gives better background suppression than would be obtained by adding the kinematic input variables directly to the BDT. 

In the \dilepton\ final state,
approximately ten optimal kinematic variables from the analysis objects and their combinations were selected for each SR,
independently for the low-mass region ($\mhp \leq 600$~\GeV) and the high-mass region ($\mhp > 600$~\GeV).
For the high-mass region, the most important variables are the scalar sum of the \pt\ of all jets and leptons, \Htall, and the transverse momentum of the jet pair with maximum \pt. For the low-mass region, the smallest invariant mass formed by two $b$-tagged jets and the smallest invariant mass formed by a lepton and a $b$-tagged jet, are among the most important variables.

\begin{figure}[htb]
  \begin{center}
\subfloat[]{ \includegraphics[width=0.45\textwidth]{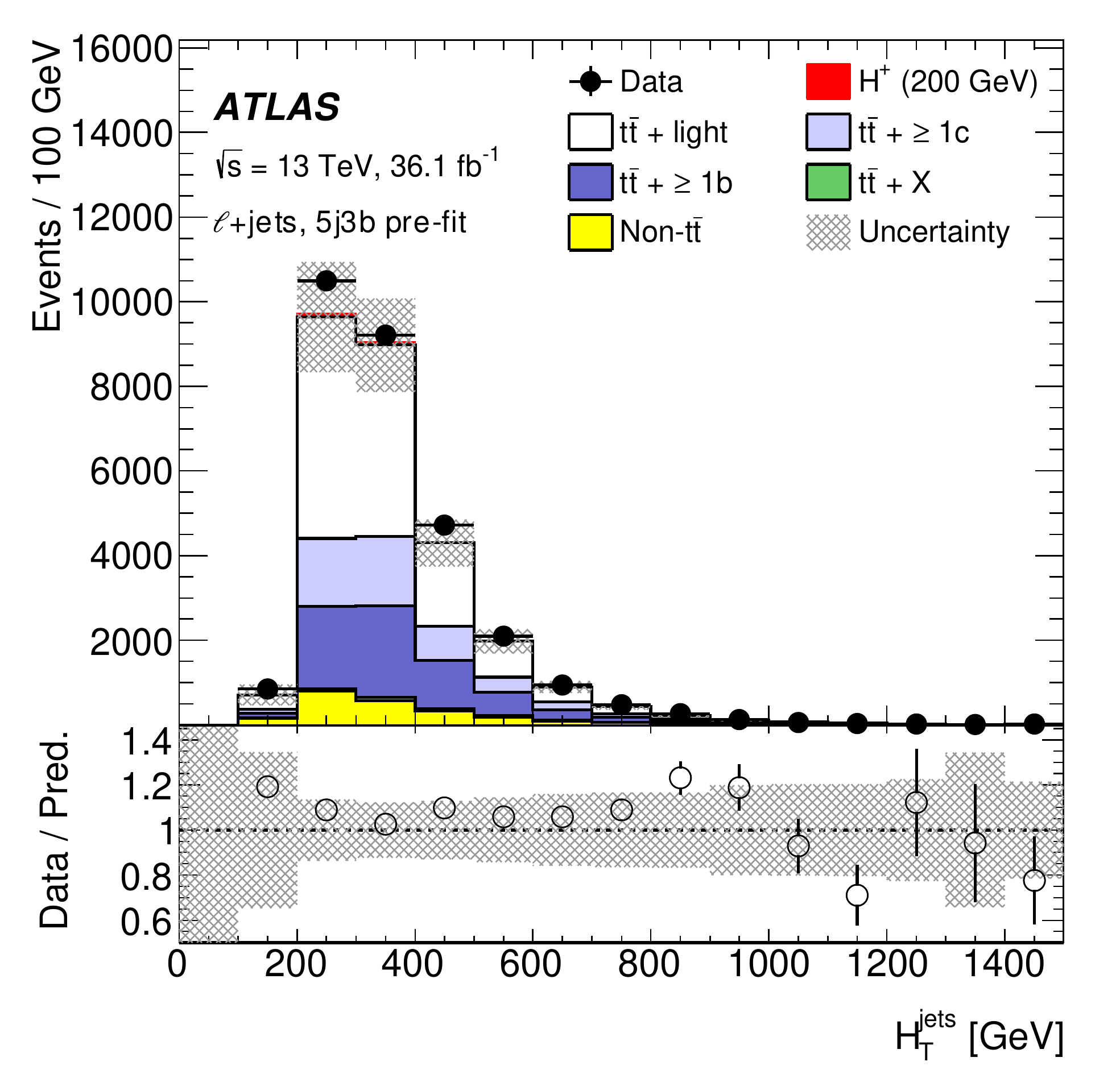} }
\subfloat[]{ \includegraphics[width=0.45\textwidth]{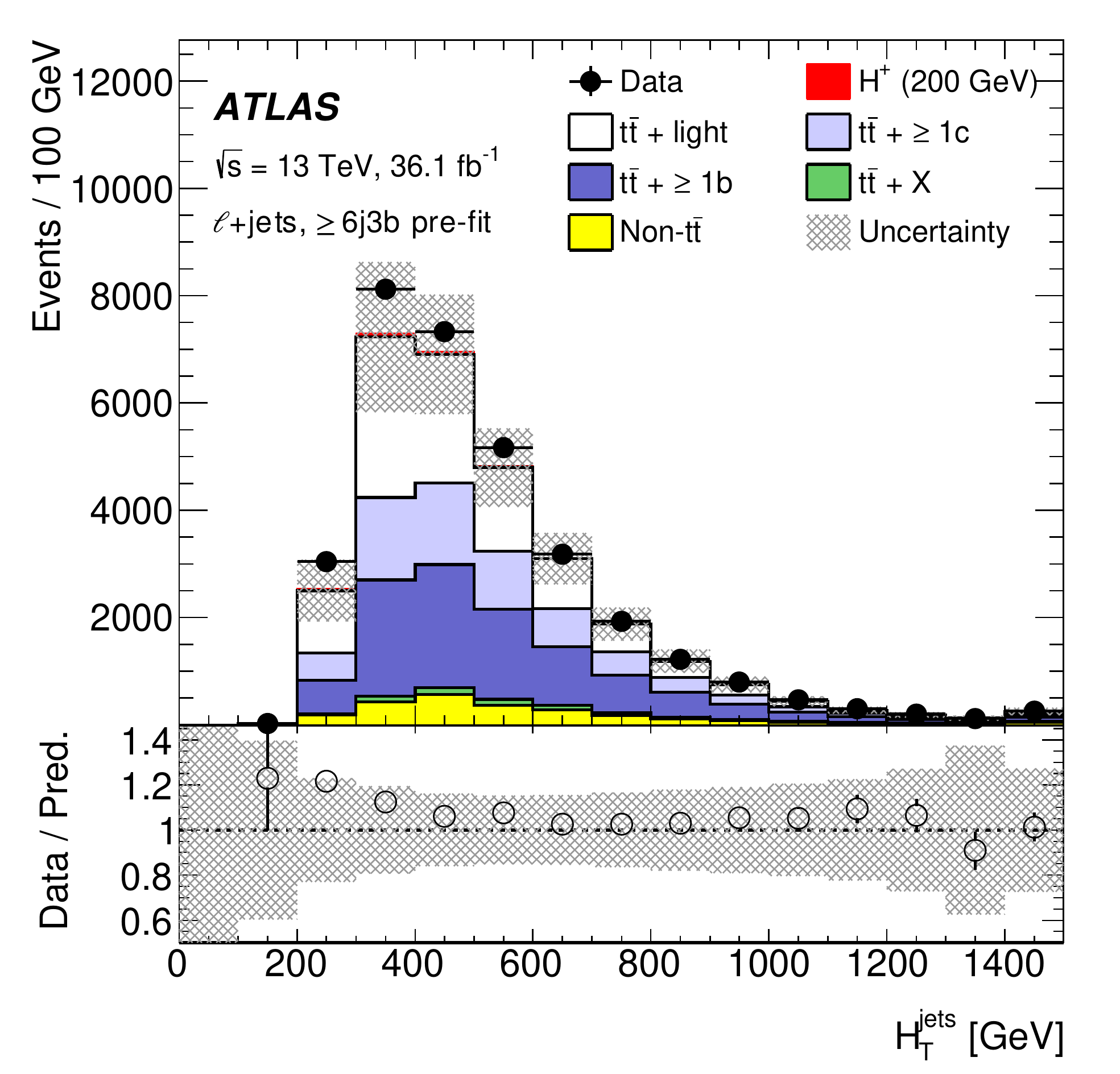} }\\
\subfloat[]{ \includegraphics[width=0.45\textwidth]{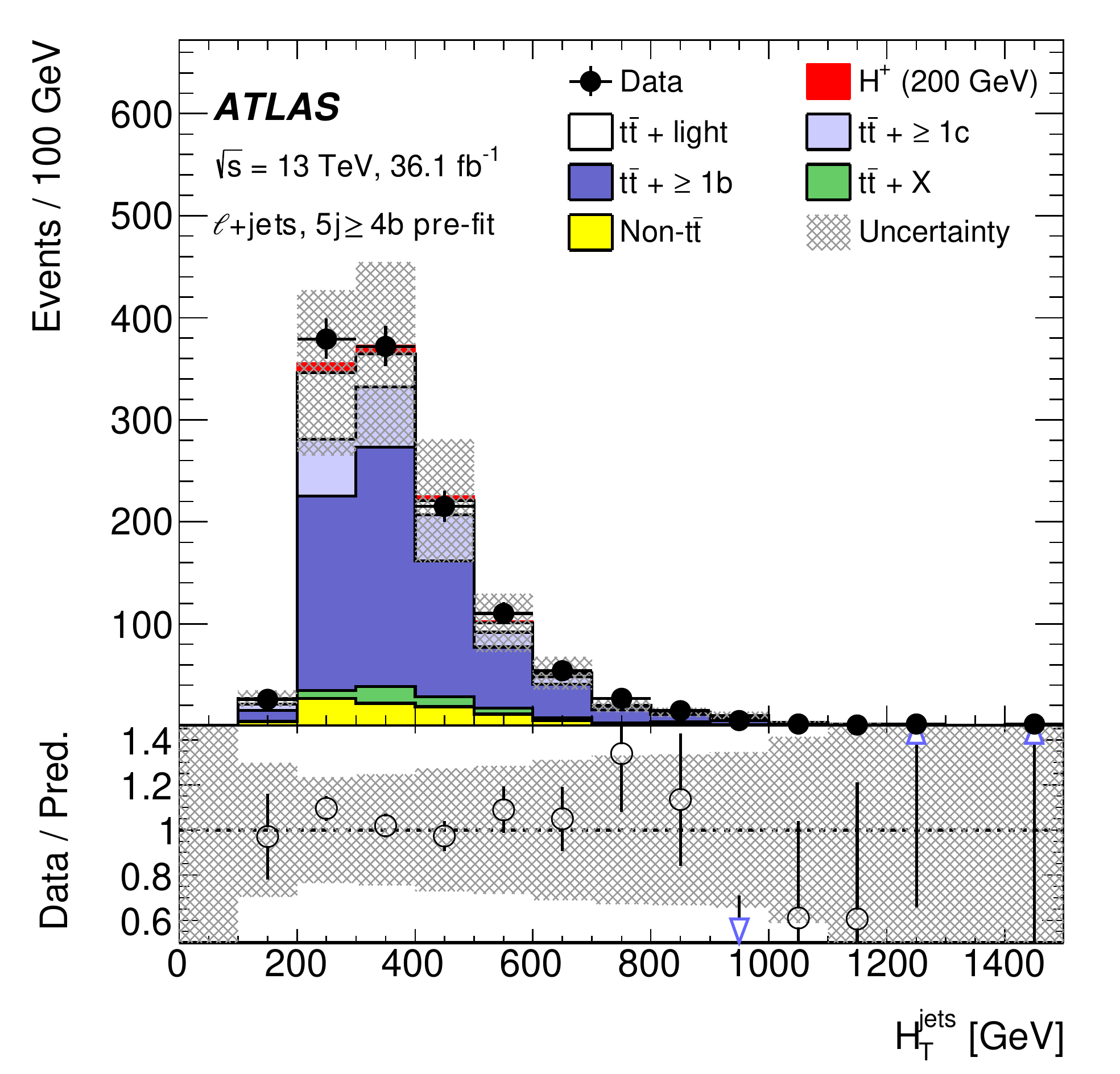} } 
\subfloat[]{ \includegraphics[width=0.45\textwidth]{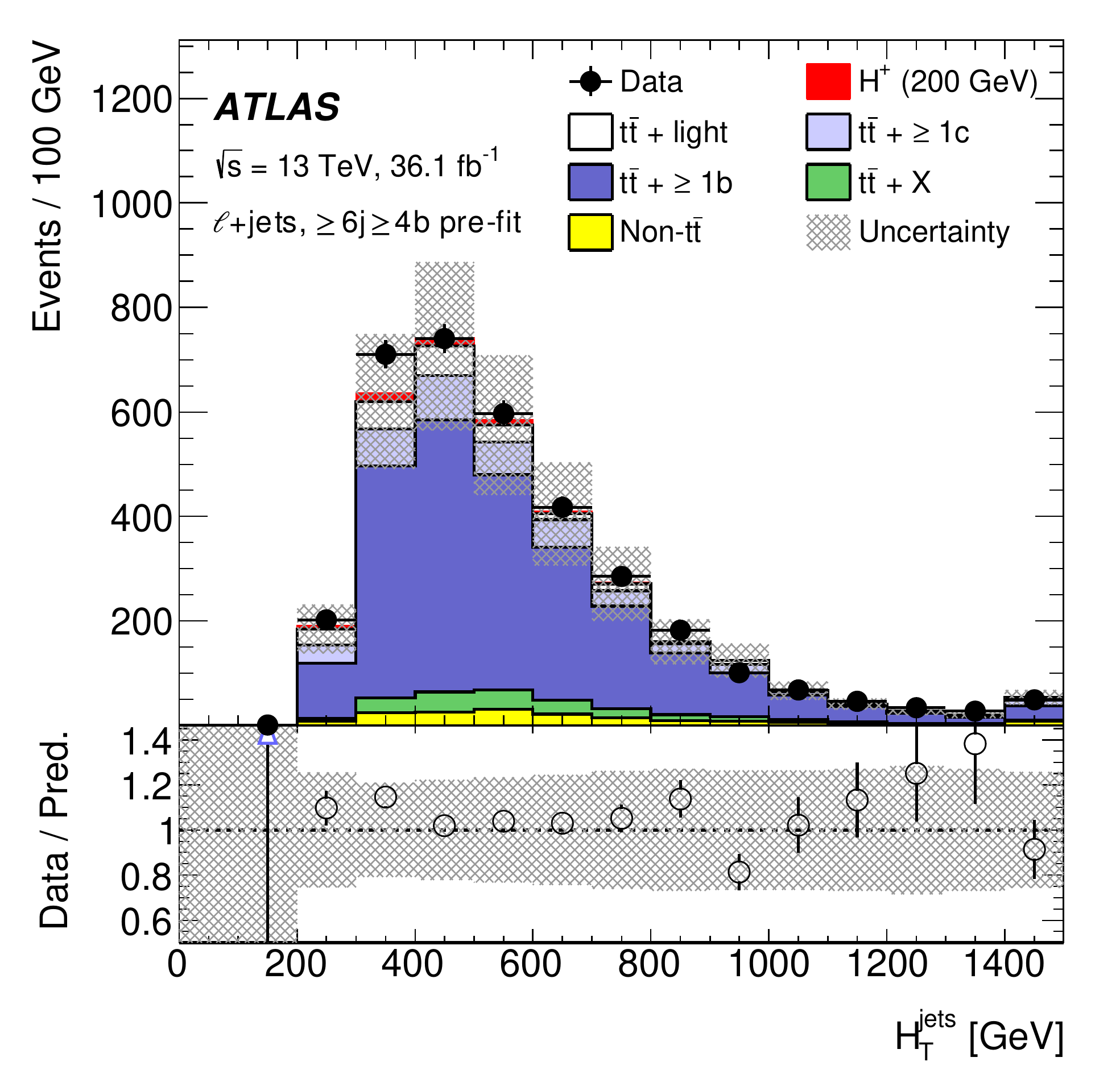} } \\
\end{center}
\caption {
Distributions of the \Htjets\ variable before the fit to the data in the four SRs of the \ljets\ channel:
(a) \FiveJexThreeBex, 
(b) \SixJinThreeBex, 
(c) \FiveJexFourBin, 
(d) \SixJinFourBin. 
Each background process is normalised according to its cross-section and the normalisation of the \ttbin\ and \ttcin\ backgrounds corresponds to the prediction from \powheg+\pythia8 for the fraction of each of these components relative to the total $\ttbar$ prediction.
The $\ttbar + X$ includes contributions from \ttW, \ttZ\ and \ttH.
In addition, the expectation for a 200~\GeV\ signal is shown for a cross-section times branching ratio of 1~pb.
The lower panels display the ratio of the data to the total prediction.
The hatched bands show the pre-fit uncertainties.
The level of agreement is improved post-fit due to the adjustment of the normalisation of the \ttgeb\ and \ttgec\ backgrounds and the other nuisance parameters by the fit.
\label{fig:HTSingleLeptonPre200} }
\end{figure}

\begin{figure}[htb]
  \begin{center}
\subfloat[]{ \includegraphics[width=0.45\textwidth]{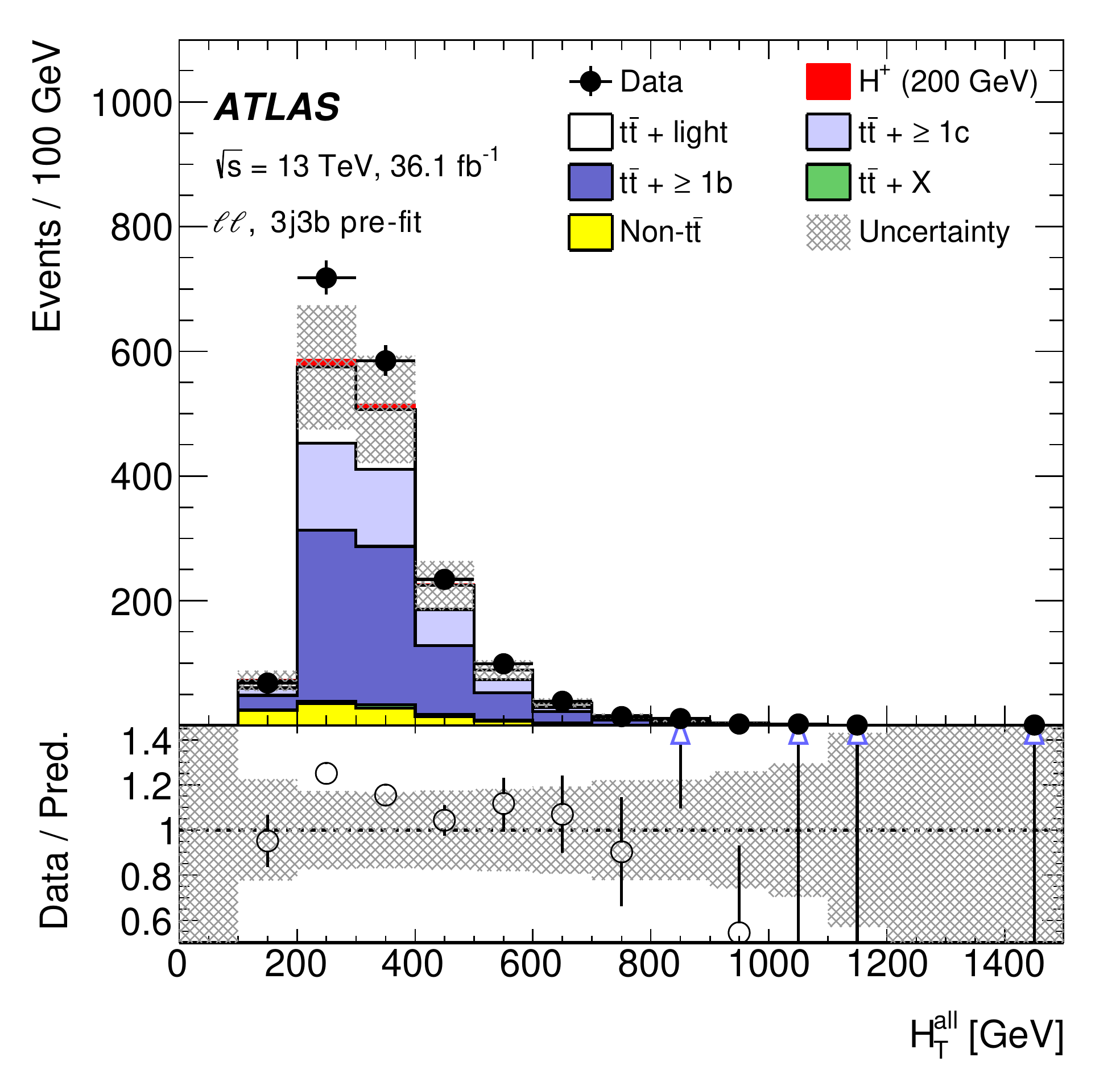} }
\subfloat[]{ \includegraphics[width=0.45\textwidth]{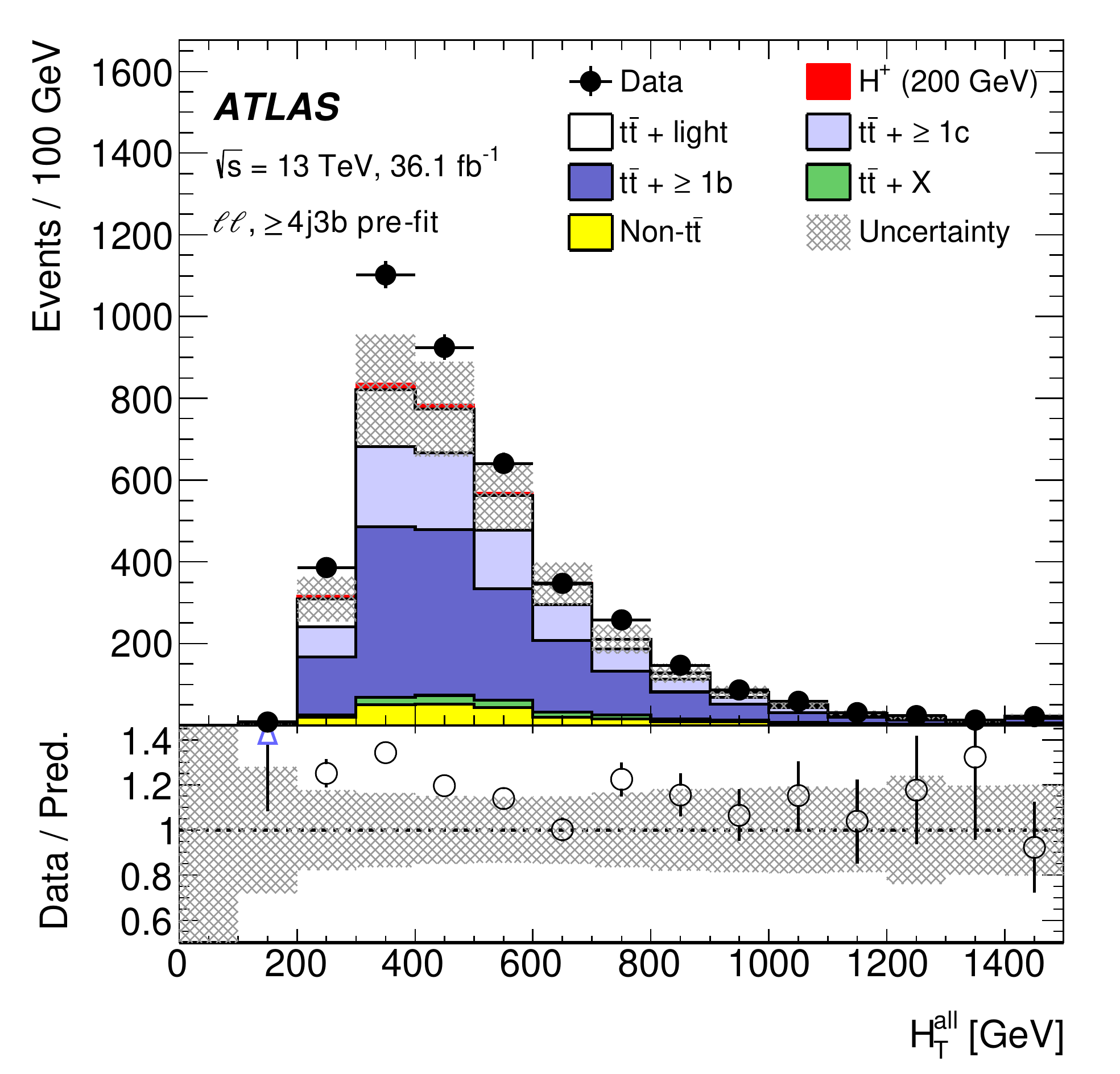} }\\
\subfloat[]{ \includegraphics[width=0.45\textwidth]{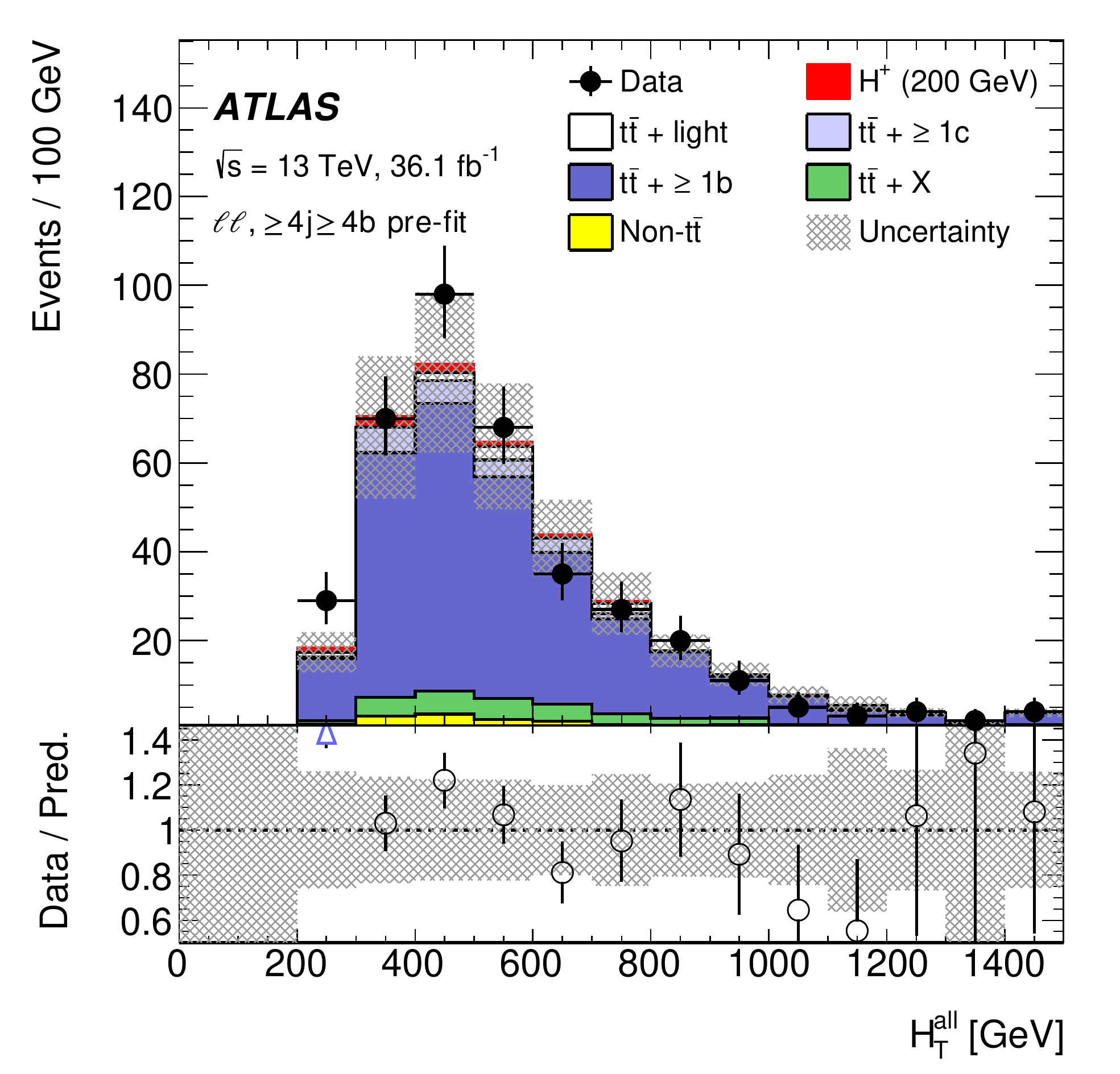} } \\
\end{center}
\caption {
Distributions of the \Htall\ variable before the fit to the data in the three SRs of the \dilepton\ channel:
(a) \ThreeJexThreeBex,
(b) \FourJinThreeBex\ and
(c) \FourJinFourBin.
Each background process is normalised according to its cross-section and the normalisation of the \ttbin\ and \ttcin\ backgrounds corresponds to the prediction from \powheg+\pythia8 for the fraction of each of these components relative to the total $\ttbar$ prediction.
The $\ttbar + X$ includes contributions from \ttW, \ttZ\ and \ttH.
In addition, the expectation for a 200~\GeV\ signal is shown for a cross-section times branching ratio of 1~pb.
The lower panels display the ratio of the data to the total prediction.
The hatched bands show the pre-fit uncertainties.
The level of agreement is improved post-fit due to the adjustment of the normalisation of the \ttgeb\ and \ttgec\ backgrounds and the other nuisance parameters by the fit.
\label{fig:HTDiLeptonPre200} }
\end{figure}

\begin{figure}[htbp]
  \begin{center}
    \subfloat[]{ \includegraphics[width=0.45\textwidth]{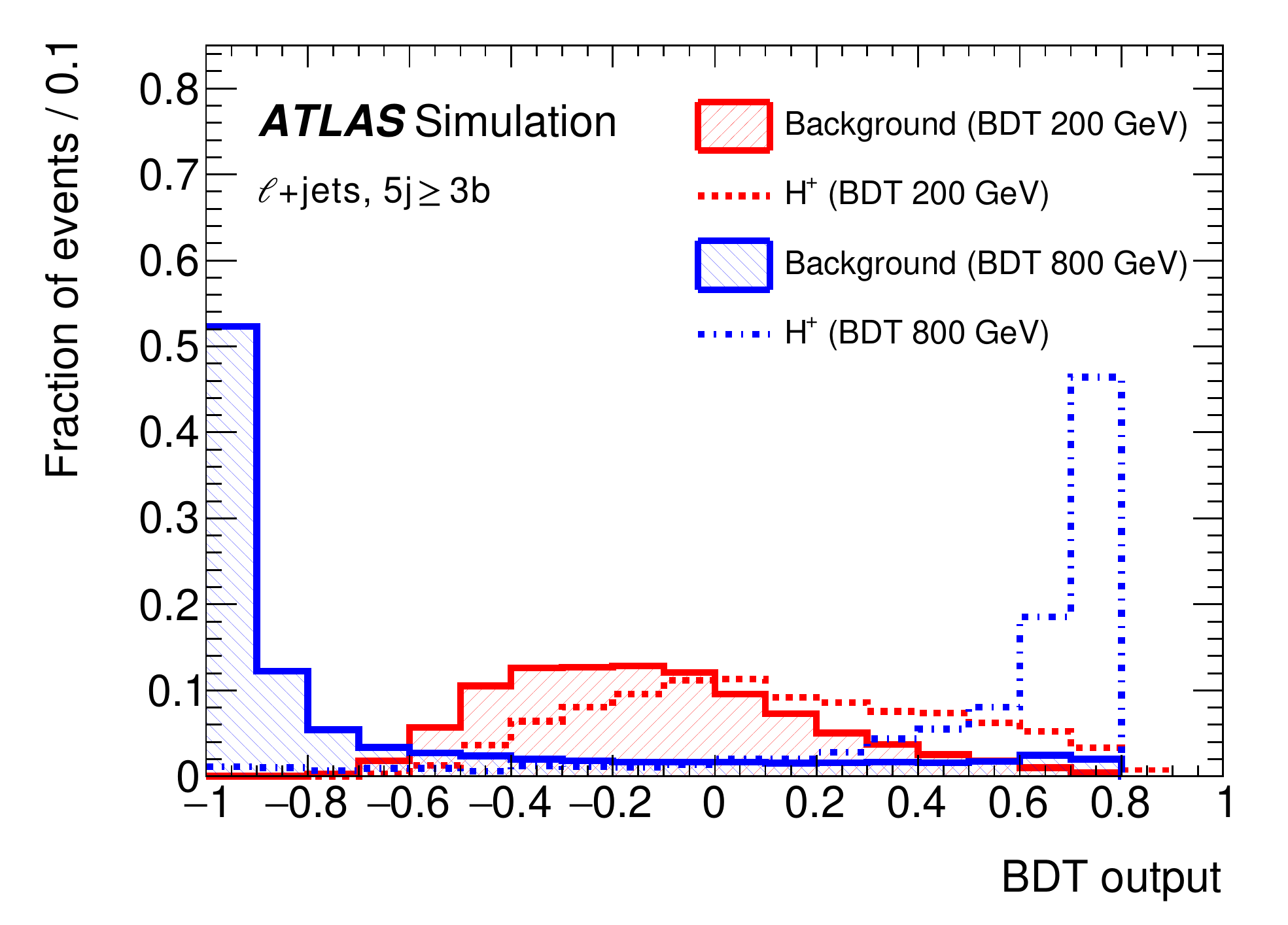} }
    \subfloat[]{ \includegraphics[width=0.45\textwidth]{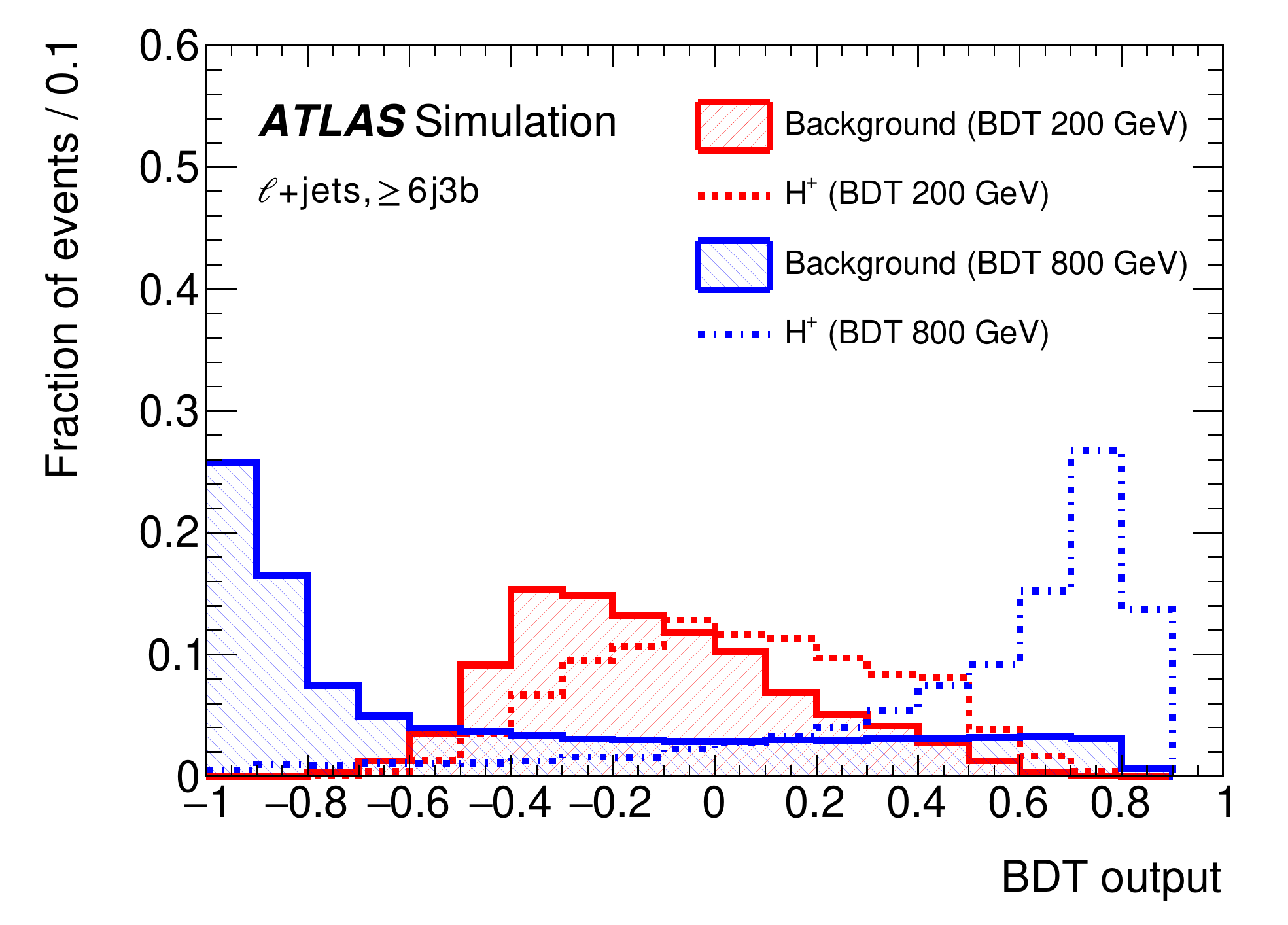} }\\
    \subfloat[]{ \includegraphics[width=0.45\textwidth]{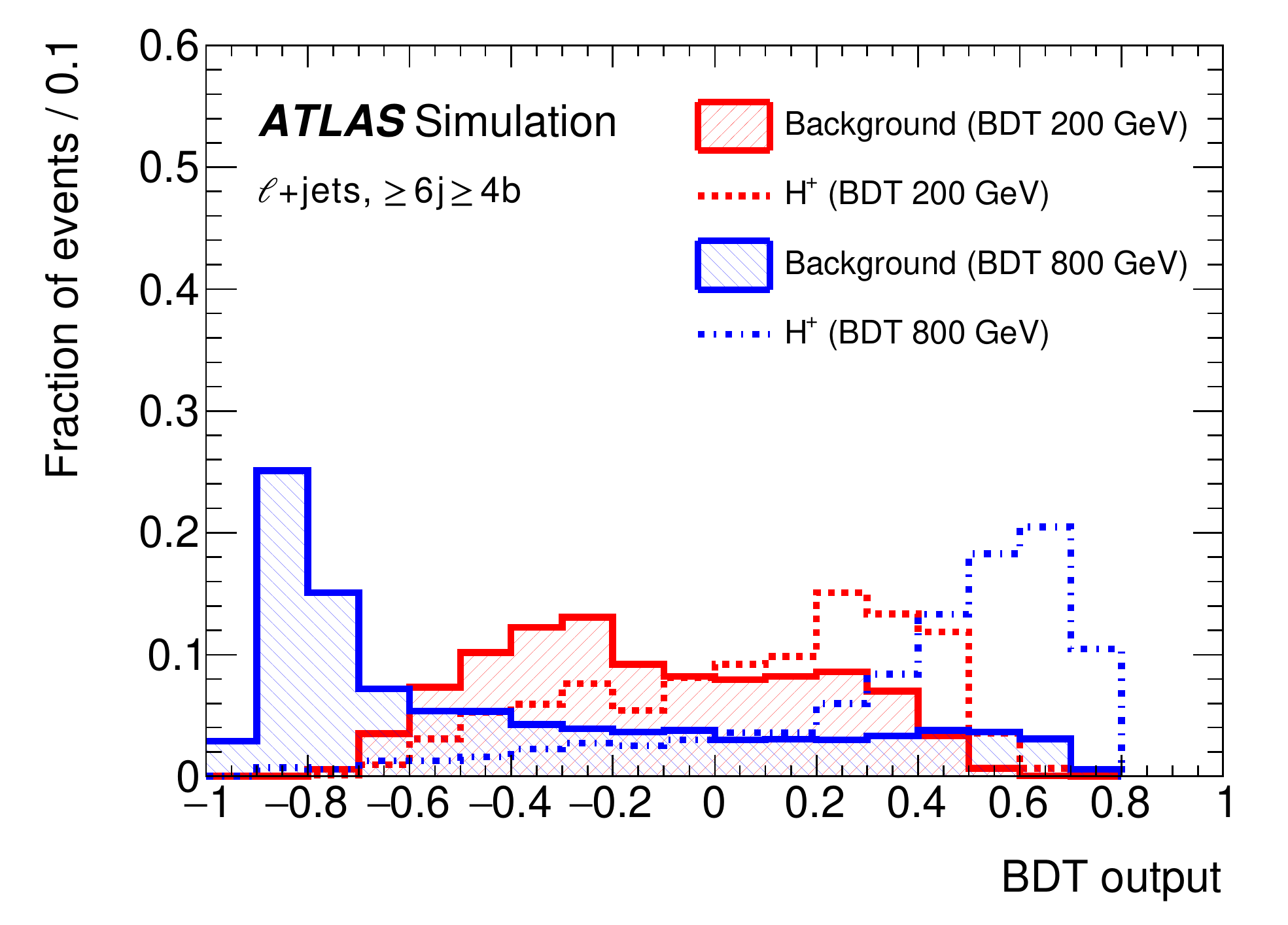} }
    \subfloat[]{ \includegraphics[width=0.45\textwidth]{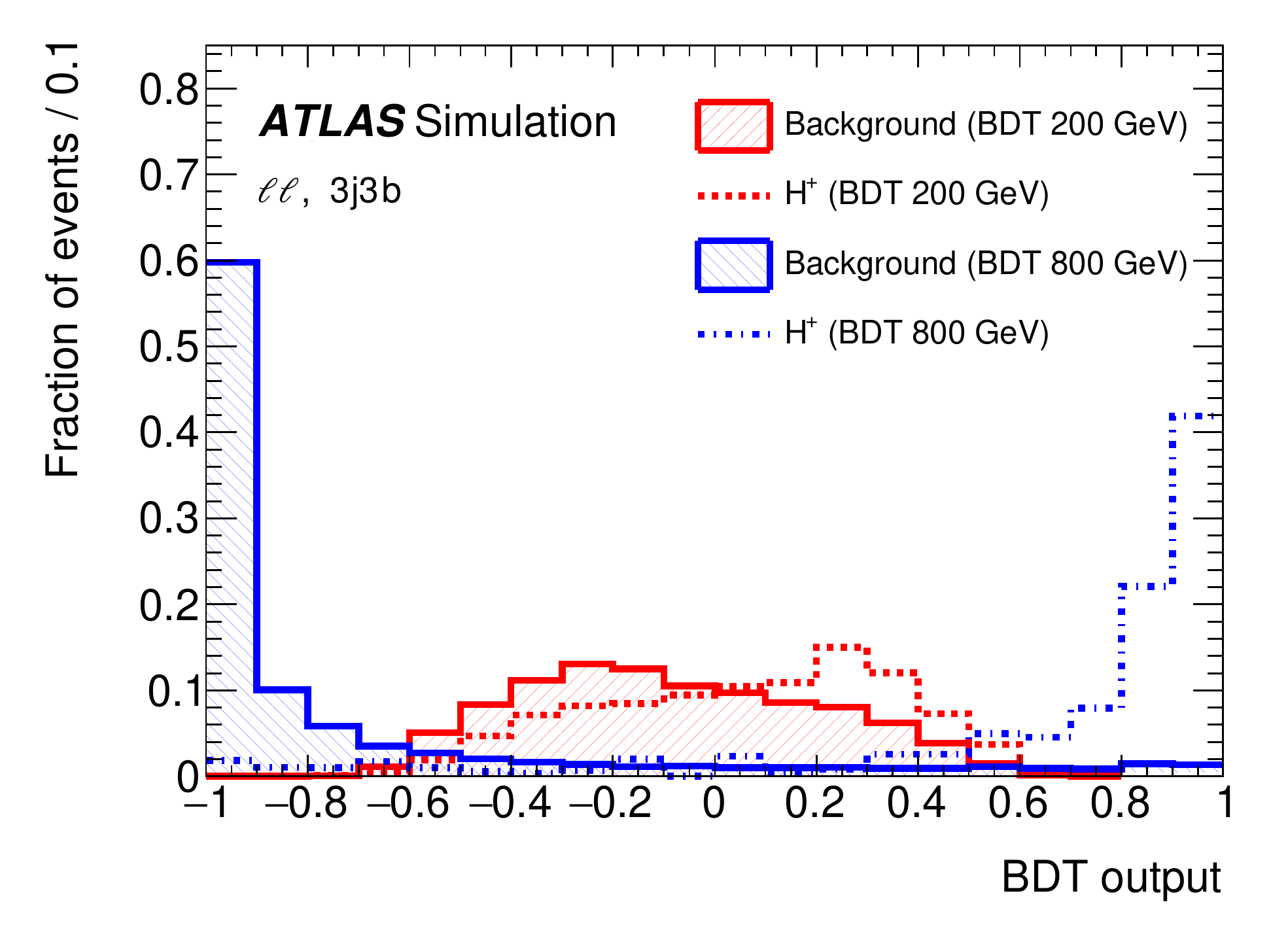} }\\
    \subfloat[]{ \includegraphics[width=0.45\textwidth]{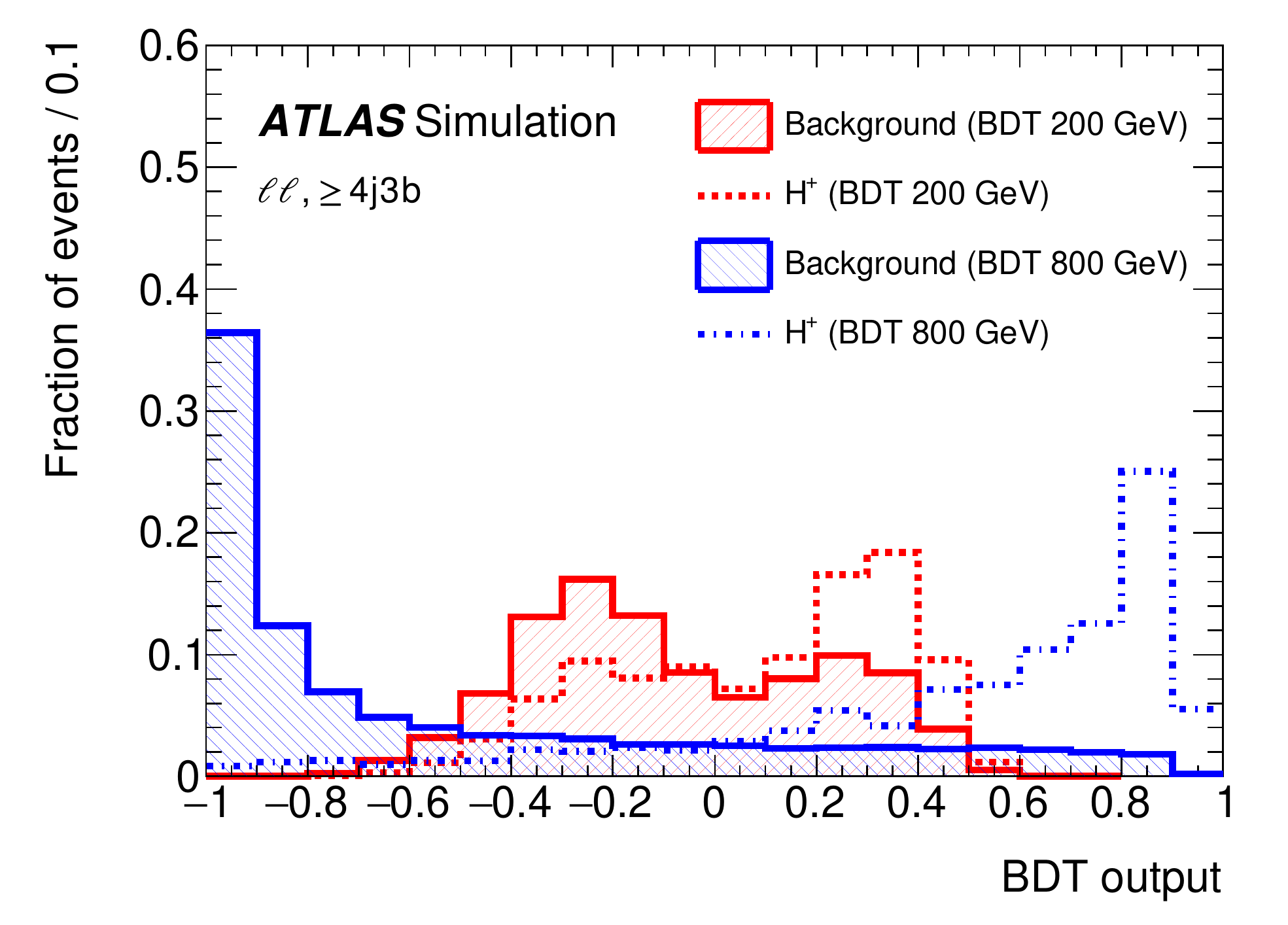} }
    \subfloat[]{ \includegraphics[width=0.45\textwidth]{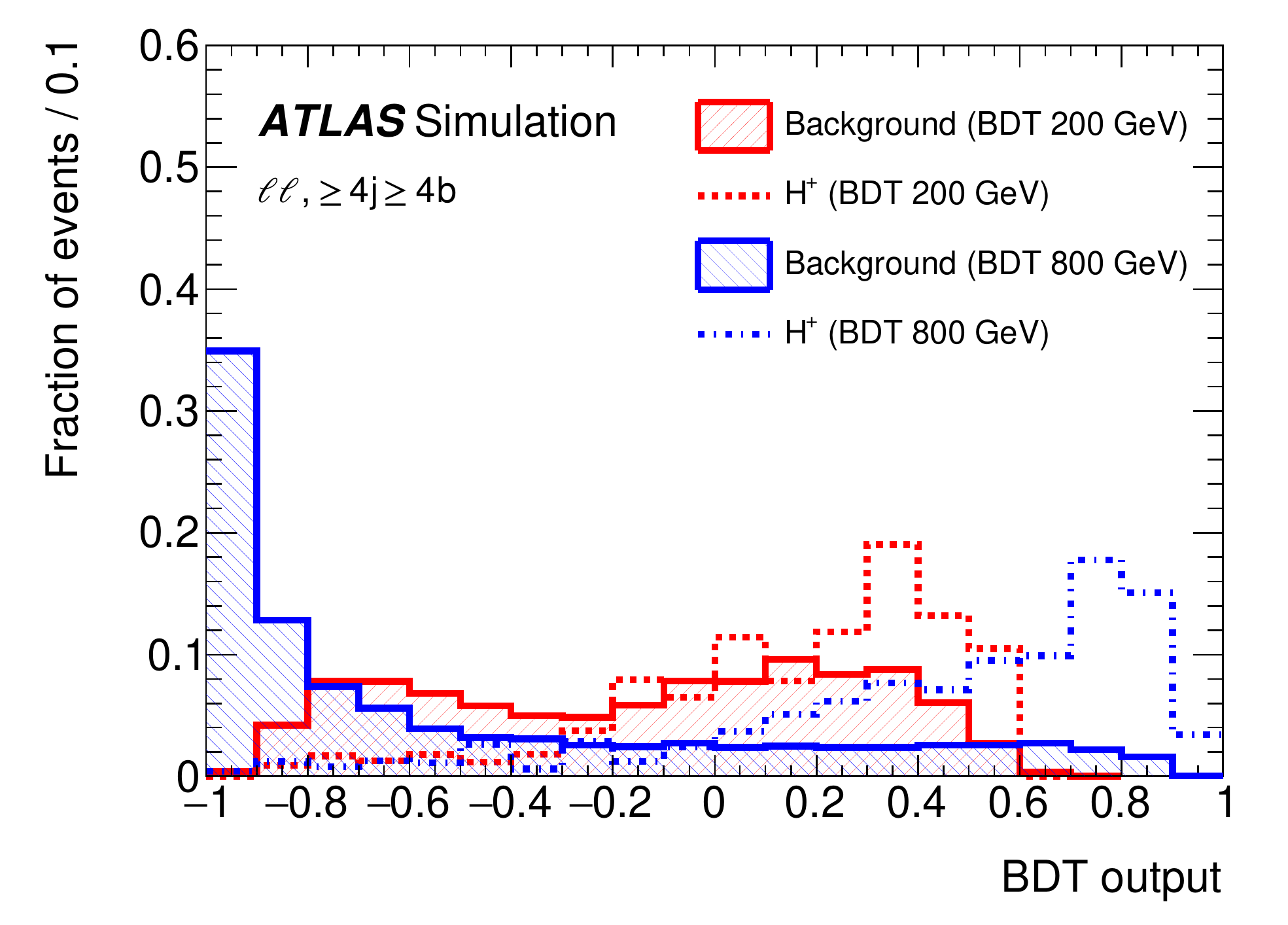} }\\
  \end{center}
\caption{
The expected output distributions of the BDTs employed for \hplus\ masses of 200~\GeV\ and 800~\GeV\ for SM backgrounds
and \hplus\ signal in the three \ljets\ and the three \dilepton\ SRs used in the BDT training:
(a) \ljets\ final state, \FiveJexThreeBin, 
(b) \ljets\ final state, \SixJinThreeBex, 
(c) \ljets\ final state, \SixJinFourBin, 
(d) \dilepton\ final state, \ThreeJexThreeBex,
(e) \dilepton\ final state, \FourJinThreeBex\ and
(f) \dilepton\ final state, \FourJinFourBin. 
All distributions are normalised to unity.
\label{fig:BDT-signal-back}
}
\end{figure}

All BDT input variables in the \ljets\ and \dilepton\ final states are listed in the Appendix.
In most regions, the distributions show a reasonable level of agreement between simulation and data within the systematic and statistical uncertainties before the fit to the data (pre-fit).
As examples, Figures \ref{fig:HTSingleLeptonPre200} and \ref{fig:HTDiLeptonPre200} show the distribution of the observed and pre-fit expected event yields for \Htjets\ in the \ljets\ channel and \Htall\ in the \dilepton\ channel.
Figure~\ref{fig:BDT-signal-back} shows the expected BDT output distributions, normalised to unity, for selected \hplus\ signal samples and the background processes in the SRs.

\clearpage

%% file: systematics.tex
\clearpage
\section{Systematic uncertainties}
\label{sec:systematics}

Systematic uncertainties from various sources affect this search, such as uncertainties in 
the luminosity measurement, the reconstruction and calibration of physics objects, in particular $b$-tagged jets, and the modelling of the signal and background processes. Uncertainties can either modify the normalisation of the signal and background processes, change the shape of the final distributions, or both. The experimental uncertainties were obtained from dedicated analyses detailed in the corresponding references. The uncertainties related to this analysis are described in this section. For a precise treatment, the uncertainties are split into several components as explained in the following. The exact number of components for each category is listed in Table~\ref{tab:SystSummary}.
The most important uncertainties are related to jet flavour tagging, background modelling, jet energy scale and resolution and the limited number of events in the simulation samples. The impact of all systematic uncertainties is listed in Table~\ref{tab:systematics} in Section~\ref{sec:result}.

\begin{table}[htp!]
\caption{List of systematic uncertainties considered. The details of the systematic uncertainties are described in Section~\ref{sec:systematics}. `N' indicates that the uncertainty is taken as normalisation-only for all processes and channels affected, while `NS' means that the uncertainty applies to both normalisation and shape. The systematic uncertainties are split into several components for a more accurate treatment. \newline
Flavour-tagging uncertainties marked (*) are different for the two sets of calibrations: the step-wise efficiency calibration for $\mhp \leq 300$~\GeV, and the 70\% efficiency point calibration elsewhere. 
\label{tab:SystSummary} } 
\centering
\begin{tabular}{ll rrr}
\textbf{Systematic uncertainty}                       & Type & \multicolumn{3}{c}{Number of components} \\
\hline \hline
Luminosity                                            & N    &&   1 &\\
Pile-up                                               & NS   &&   1 &\\
\hline 
Electron reconstruction                               & NS   &&   6 &\\
Muon reconstruction                                   & NS   &&  13 &\\
Jet and \MET reconstruction                           & NS   &&  28 &\\
\hline
Flavour tagging, 70\% efficiency calibration (*)      & NS   &&  27 &\\
Flavour tagging, step-wise efficiency calibration (*) & NS   && 126 &\\
\hline 
Signal QCD scale and PDF                              & NS   &&  31 &\\
\hline 
Background modelling, \ttbar + jets                   & NS   &&  29 &\\
Background modelling, other top                       & NS   &&  25 &\\
Background modelling, non-top (\ljets\ final state)   & N    &&  13 &\\
Background modelling, non-top (\dilepton\ final state)& N    &&   4 &\\
\end{tabular}
\end{table}

The combined uncertainty in the integrated luminosity for the data collected in 2015 and 2016 is \lumireluncert, and it is applied as a normalisation uncertainty for all processes estimated using simulation. It is derived, following a methodology similar to that detailed in Ref.~\cite{Aaboud:2016hhf}, from a preliminary calibration of the luminosity scale using $x$--$y$ beam-separation scans performed in August 2015 and May 2016.
A variation in the pile-up reweighting of MC events is included to cover the uncertainty in the ratio of the predicted and measured inelastic cross-sections in the fiducial volume defined by $M_X > 13\,\GeV$ where $M_X$ is the mass of the hadronic system~\cite{Aaboud:2016mmw}.

Uncertainties associated with charged leptons arise from the trigger selection, the object reconstruction, the identification, and the isolation criteria, as well as the lepton momentum scale and resolution. 
These are estimated by comparing  $Z\to \ell^+\ell^-$ ($\ell=e,\, \mu$) events in data and simulation~\cite{Aad:2016jkr, ATLAS-CONF-2016-024}. 
Correction factors are applied to the simulation to better model the efficiencies observed in data. 
The charged-lepton uncertainties have a small impact on the analysis. 

Uncertainties associated with jets arise from the jet reconstruction and identification efficiencies related to the JES and jet energy resolution, and on the Jet Vertex Tagger efficiency~\cite{PERF-2011-04}. 
The JES-related uncertainties contain 23 components that are treated as statistically independent and uncorrelatd. The JES and its uncertainty were derived by combining information from test-beam data, LHC collision data (\emph{in situ} techniques) and simulation~\cite{PERF-2016-04}. The many sources of uncertainties related to the in situ calibration using $Z$+jets, $\gamma$+jets and multi-jet data were reduced to eight uncorrelated components through an eigen-decomposition. Other components are relativ to jet flavour, pile-up corrections, $\eta$-dependence and high-$p_T$ jets.

In the reconstruction of quantities used for the BDT, \met\ is used. The \met\ calculation depends on the reconstruction of leptons and jets. The uncertainties associated with these objects are therefore propagated to the \met\ uncertainty estimation. Uncertainties due to soft objects (not included in the calculation of the leptons and jets) are also considered~\cite{PERF-2016-07}. 

Differences between data and simulation in the $b$-tagging efficiency for $b$-jets, $c$-jets and light jets are taken into account using correction factors. 
For $b$-jets, the corrections are derived from \ttbar\ events with final states containing two leptons, and the corrections are consistent with unity within uncertainties at the level of a few percent over most of the jet \pt\ range.
The mis-tag rate for $c$-jets is also measured in \ttbar\ events, identifying hadronic decays of $W$ bosons including $c$-jets.
For light jets, the mis-tag rate is measured in multi-jet events using jets containing secondary vertices and tracks with impact parameters consistent with a negative lifetime.
Systematic uncertainties affecting the correction factors are derived in the \pt\ and $\eta$ bins used for extracting the correction factors. They are transformed into uncorrelated components using an eigenvector decomposition, taking into account the bin-to-bin correlations~\cite{Aad:2015ydr,ATL-PHYS-PUB-2016-012,ATLAS-CONF-2018-001}.
For $\mhp > 300\,\GeV$, corrections corresponding to the fixed working point of 70\% efficiency are used and a total of 6, 3 and 16 independent uncorrelated eigen-variations are considered as systematic uncertainties for $b$-, $c$- and light jets, respectively.
For $\mhp \leq 300\,\GeV$, corrections for the step-wise efficiencies are used to support the kinematic discriminant $D$ and the number of eigen-variations is increased by a factor of five to account for the five $b$-tagging efficiency bins. 
In addition, uncertainties due to tagging the hadronic decays of $\tau$-leptons as $b$-jets are considered. 
For $\mhp>300\,\GeV$, an additional uncertainty is included due to the extrapolation of scale factors for jets with $\pt>300$ \GeV, beyond the kinematic reach of the data calibration samples used~\cite{ATL-PHYS-PUB-2016-012}. 

The uncertainty due to different scale choices in the \Hp\ signal is estimated by varying the renormalisation and factorisation scales up and down by a factor of two. 
The uncertainty ranges from 7\% at low masses to 15\% at masses above 1300~\GeV\ for the \ljets\ final state, and from 12\% to 16.5\% for the \dilepton\ final state.  
The PDF uncertainty in the modelling is estimated using the PDF4LHC15\_30 PDF set~\cite{Butterworth:2015oua},
which is based on a combination of the CT14~\cite{ct14}, MMHT14~\cite{mmht14} and NNPDF3.0~\cite{Ball:2014uwa} PDF sets
and contains 30 components obtained using the Hessian reduction method~\cite{hessian-reduction-1,hessian-reduction-2,mc-comb-method}.

The modelling of the \ttbar + jets background is one of the largest sources of uncertainty in the analysis and many different components are considered.
The uncertainty in the inclusive \ttbar\ production cross-section at NNLO+NNLL~\cite{Czakon:2011xx} is $6\%$, 
including effects from varying the factorisation and renormalisation scales, the PDF, the QCD coupling constant $\alpha_{\mathrm{s}}$, and the top quark mass.
Due to the large difference between the 4FS prediction and the various 5FS predictions for the \ttbbb\ process, an additional 50\% normalisation uncertainty is assigned to this background. 

The uncertainty due to the choice of NLO generator is derived by comparing the nominal \powheg\ sample with a sample generated using \SHERPAV{2.2.1} with a 5FS PDF. 
A \powheg\ sample with the same settings as in the nominal \powheg+\pythia8 sample, but using \herwig7~\cite{Bahr:2008pv,Bellm:2015jjp} for parton showering, is used to assess the uncertainty due to the choice of parton shower and hadronisation model. 
Furthermore, the uncertainty due to the modelling of initial- and final-state radiation is evaluated with two different \powheg+\pythia8 samples in which the radiation is increased or decreased by halving or doubling the renormalisation and factorisation scales in addition to simultaneous changes to the \hdamp parameter and the A14 tune parameters~\cite{ATL-PHYS-PUB-2016-004}. 

For the \ttbin\ background, an additional uncertainty is assigned by comparing the predictions from \powheg+\pythia8 and \SHERPA with 4FS. This takes into account the difference between a 5FS inclusive \ttbar\ prediction at NLO and a 4FS NLO $t\bar{t}b\bar{b}$ prediction.
For the \ttcin\ background, an additional uncertainty is derived by comparing a \mg5\_\amc\ sample that is interfaced to \HERWIGpp~\cite{Bahr:2008pv} with the nominal event sample. 
In this \mg5\_\amc\ event sample, a three-flavour scheme is employed and the $t\bar{t}c\bar{c}$ process is generated at the matrix element level~\cite{ATL-PHYS-PUB-2016-011} using the CT10F3 PDF set, while in the nominal sample the charm jets are primarily produced in the parton shower. 
All of these uncertainties, with the exception of the inclusive and \ttbbb\ cross-sections, are considered to be uncorrelated amongst the \ttbin, \ttcin, and \ttbar + light samples. 
For the modelling of the \ttbin\ backgrounds, the alternative samples are reweighted to the NLO prediction of $\ttbar\bbbar$ from \SHERPA before the uncertainty is evaluated.

In addition, uncertainties due to the reweighting to the \SHERPA NLO prediction of $\ttbar\bbbar$ are considered.
For these uncertainties, the \ttbin\ is reweighted to different \SHERPA predictions with modified scale parameters, in particular where the renormalisation scale is varied up and down by a factor of two,
where the functional form of the resummation scale is changed to $\mu_\mathrm{CMMPS}$
and where a global scale choice $\muQ=\muR=\muF=\mu_\mathrm{CMMPS}$ is used. 
Two alternative
PDF sets, MSTW2008NLO~\cite{Martin:2009iq} and NNPDF2.3NLO~\cite{Ball:2014uwa}, are used, and uncertainties in the underlying event and parton shower are estimated from samples with an alternative set of tuned parameters for the underlying event and an alternative shower recoil scheme.
Due to the absence of $b$-jets from multi-parton interactions and final-state gluon radiation in the $\ttbar\bbbar$ prediction from \SHERPA, a 50\% uncertainty is assigned to the \ttb\ (MPI/FSR) category based on studies of different sets of UE tunes.
An uncertainty due to the reweighting of the leading jet \pt\ is determined by comparing a reweighted event sample with an event sample without reweighting. 
Because the reweighting changes the normalisation for jet $\pt >400$~\GeV\ by 15\%, an additional normalisation uncertainty of 15\% is applied in this region.
The reweighting factors are derived from the CR with exactly four jets and at least two $b$-tagged jets and applied to higher jet multiplicity bins.
However, the effect of this extrapolation is expected to be small and is covered by the above uncertainties.

An uncertainty of $5\%$ is assigned to the total cross-section for single top-quark production~\cite{Kidonakis:2010ux,Kidonakis:2010tc,Kidonakis:2011wy}, uncorrelated between $Wt$ and $t$-channel production. 
An additional uncertainty due to initial- and final-state radiation is estimated using samples with factorisation and renormalisation scale variations and appropriate variations of the Perugia 2012 set of tuned parameters. 
The parton showering and hadronisation modelling uncertainties in the single-top $Wt$ and $t$-channel production are estimated by comparing with samples where the parton shower generator is \HERWIGpp\ instead of \PYTHIAV{6.428}.
The uncertainty in the interference between $Wt$ and \ttbar\ production at NLO~\cite{Frixione:2008yi} is assessed by comparing the default `diagram removal' scheme with an alternative `diagram subtraction' scheme~\cite{Frixione:2008yi,PowhegWt}. 

The uncertainty arising from \ttV\ generation is estimated by comparison with samples generated with \SHERPA.
The uncertainty in the \ttV\ production cross-section is about 15\%, taken from the NLO predictions~\cite{Kardos:2011na,Lazopoulos:2008de,Campbell:2012dh,deFlorian:2016spz},
treated as uncorrelated between $t\bar{t}W$ and $t\bar{t}Z$ with PDF and QCD scale variations.

The \ttH\ modelling uncertainty is assessed through an uncertainty in the cross-section, uncorrelated between QCD ($^{+5.8}_{-9.2}$\%) and the PDFs ($\pm3.6$\%)~\cite{deFlorian:2016spz,Raitio:1978pt,Beenakker:2002nc,Dawson:2003zu,Yu:2014cka,Frixione:2015zaa}, and the modelling of the parton shower and hadronisation by comparing \pythia8\ with \HERWIGpp. 
The minor $\tH+X$ backgrounds, \tHjb\ and \WtH\ are  treated as one background and its cross-section uncertainty is 6\% due to PDF uncertainties and another 10\% due to factorisation and renormalisation scale uncertainties~\cite{deFlorian:2016spz}. 

The uncertainties from the data-driven estimation of non-prompt leptons are based on a comparison between data and the non-prompt lepton estimates in CRs. 
A 50\% uncertainty is assigned in the $\ell$+jets final state. 
In the \dilepton\ final state, where all backgrounds with one or no prompt leptons fall into this category, including $W$+jets and single top production, an uncertainty of $25\%$ is assigned. 

An uncertainty of 40\% is assumed for the $W$+jets cross-section, uncorrelated between jet bins, with an additional 30\% for $W$+HF jets, uncorrelated for two, three and more than three HF jets. These uncertainties are derived from variations of the renormalisation and factorisation scales and matching parameters in \SHERPA simulations.
An uncertainty in $Z$+jets of 35\% is applied, uncorrelated among jet bins in the \dilepton\ final state. This uncertainty accounts for both the variation of the scales and matching parameters in \SHERPA simulations and the data-driven correction factors applied to the $Z$+HF jets component.
In the \dilepton\ final state, only the $Z+$jets component is estimated separately, and the $W+$jets background is included in the estimation of the background from non-prompt leptons.

%% file: statistical.tex
\section{Statistical analysis}
\label{sec:statistical}
In order to test for the presence of an \hplus\ signal, a binned maximum-likelihood fit to the data is performed simultaneously in all categories,
and each mass hypothesis is tested separately. 
The inputs to the fit include the number of events in the CRs and the binned BDT output in the SRs. 
Two initially unconstrained fit parameters are used to model the normalisation of the \ttbin\ and \ttcin\ backgrounds. 
The procedures used to quantify the level of agreement with the background-only or background-plus-signal hypothesis and to determine exclusion limits are based on the profile likelihood ratio test  and the $\mathrm{CL}_\mathrm{s}$ method~\cite{Asimov,CLs,Junk:1999kv}.
The parameter of interest is the signal strength, $\mu$, defined as the product of the production cross-section $\sigma(pp\to \tbhplus)$ and the branching ratio $\BR(\hplustb)$.

To estimate the signal strength, a likelihood function, $\mathcal{L}(\mu, \theta)$, is constructed as the product of Poisson probability terms. 
One Poisson term is included for every CR and every bin of the BDT distribution in the SRs. 
The expected number of events in the Poisson terms is a function of $\mu$, and a set of nuisance parameters, $\theta$. 
The nuisance parameters encode effects from the normalisation of backgrounds, including two free normalisation factors for the \ttgeb\ and \ttgec\ backgrounds, the systematic uncertainties and one parameter per bin to model statistical uncertainties in the simulated samples. 
All nuisance parameters are constrained with Gaussian or log-normal terms.
There are about 170 nuisance parameters considered in the fit, the number varying slightly across the range of mass hypotheses. 

To extract the exclusion limit on $\mu = \sigma(pp \rightarrow \tbhplus) \times \BR(\hplustb)$, the following test statistic is used:
\begin{equation*}
\tilde{t}_\mu = \left\{ 
	\begin{array}{ll}
		-2 \ln \frac{ \mathcal{L}\left( \mu, \hat{\hat{\theta}}(\mu) \right) } { \mathcal{L}\left( 0, \hat{\hat\theta}(0) \right) }  &  \hat\mu < 0,\\
		-2 \ln \frac{ \mathcal{L} \left( \mu, \hat{\hat\theta}(\mu) \right) } { \mathcal{L}\left( \hat\mu, \hat\theta \right) }  &  \hat\mu \geq 0.
	\end{array}
\right. 
\end{equation*}
The values of the signal strength and nuisance parameters that maximise the likelihood function are represented by $\hat\mu$ and $\hat\theta$, respectively. 
For a given value of $\mu$, the values of the nuisance parameters that maximise the likelihood function are represented by $\hat{\hat\theta}(\mu)$.

%% file: results.tex
\section{Results}
\label{sec:result}

\begin{figure}[htb]
  \begin{center}
\subfloat[]{ \includegraphics[width=0.45\textwidth]{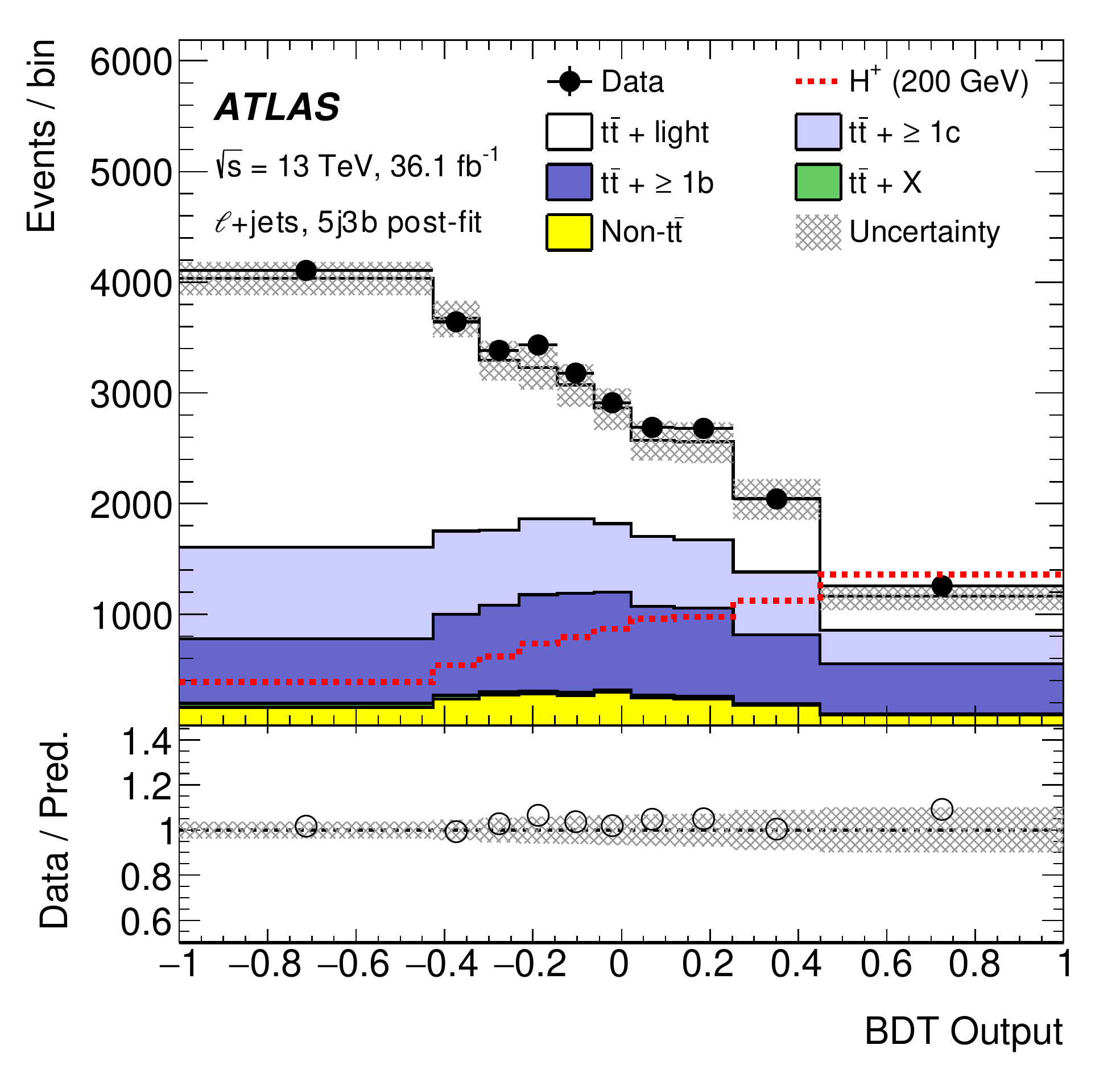} }
\subfloat[]{ \includegraphics[width=0.45\textwidth]{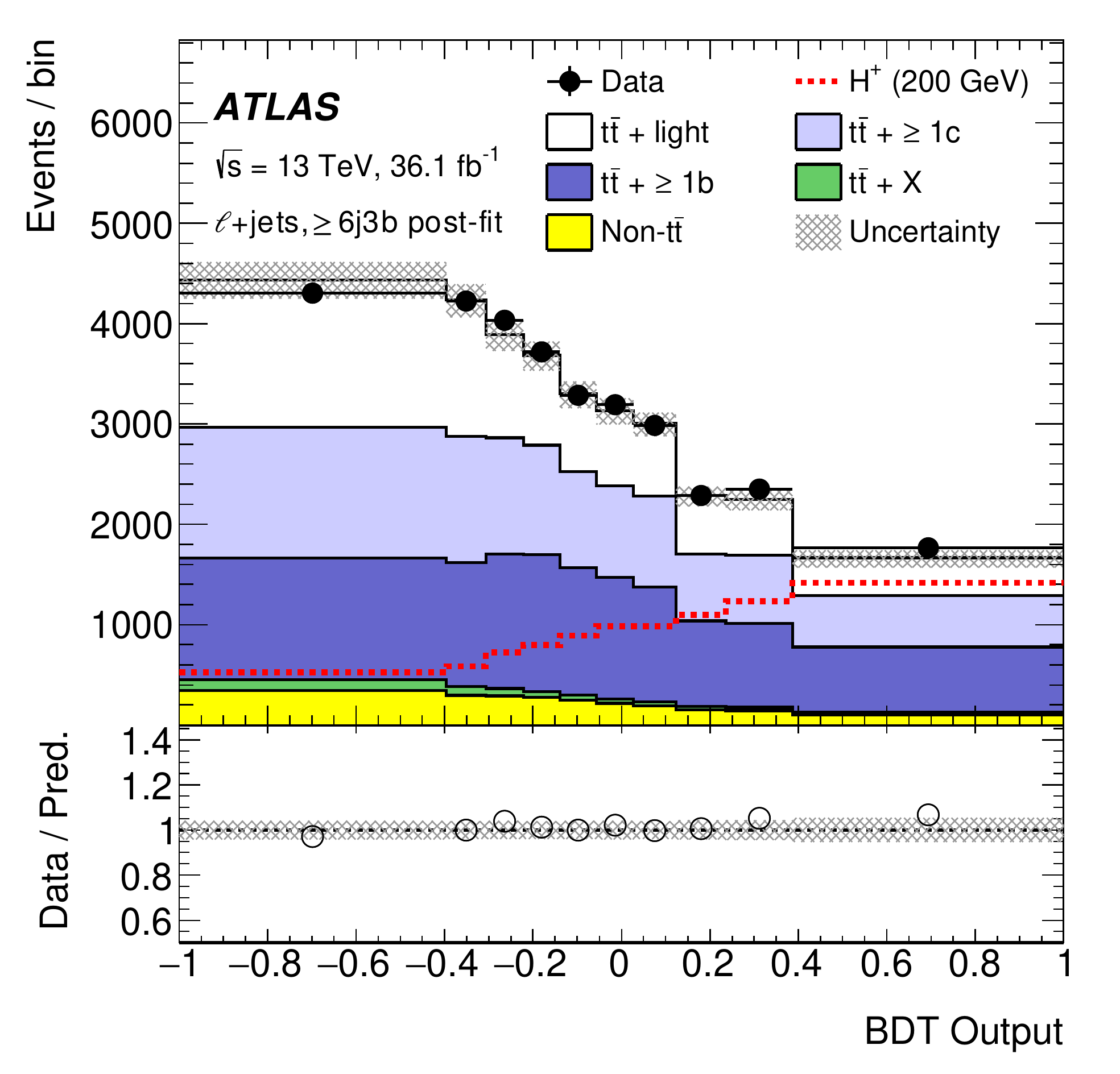} }\\
\subfloat[]{ \includegraphics[width=0.45\textwidth]{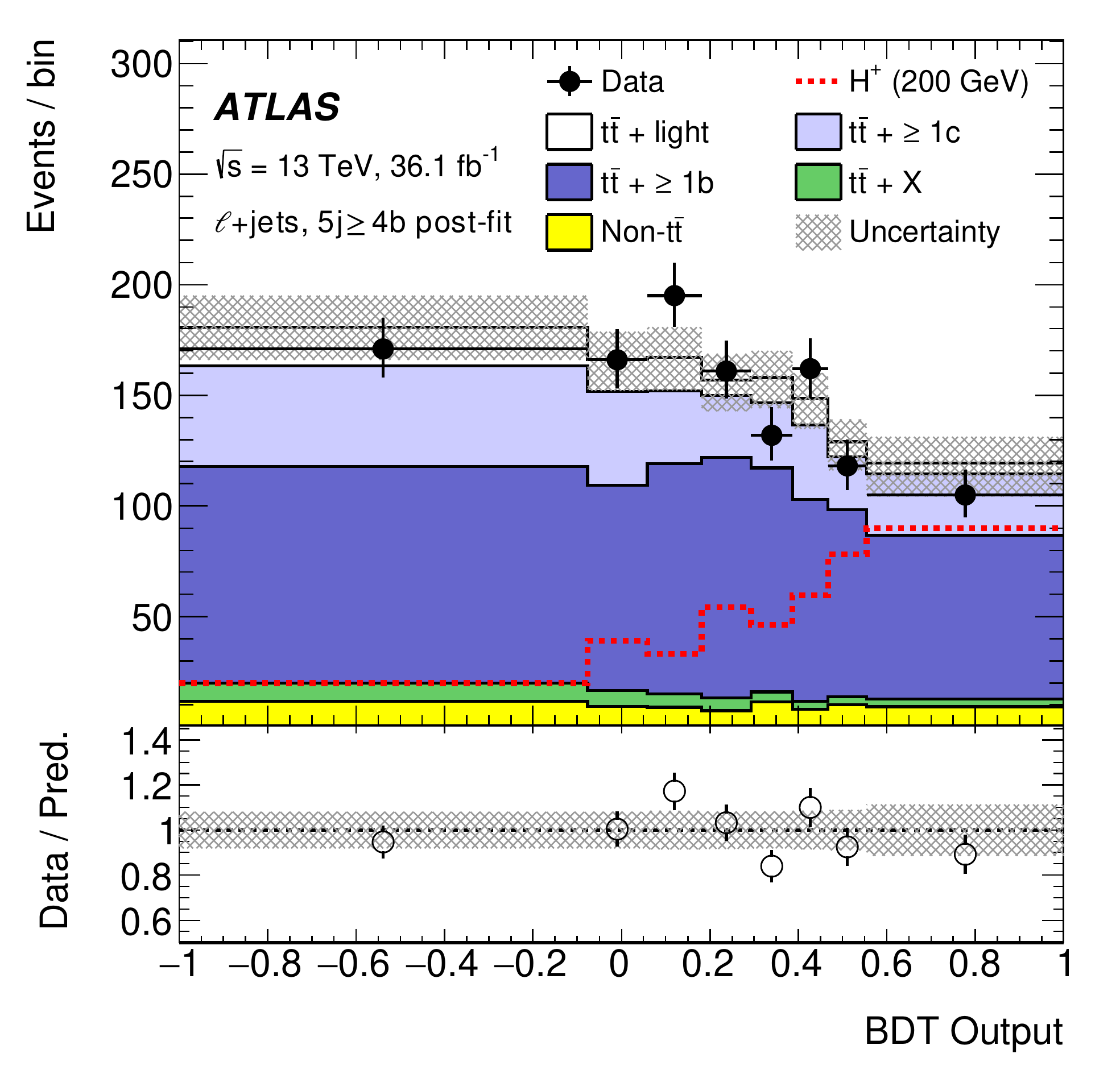} } 
\subfloat[]{ \includegraphics[width=0.45\textwidth]{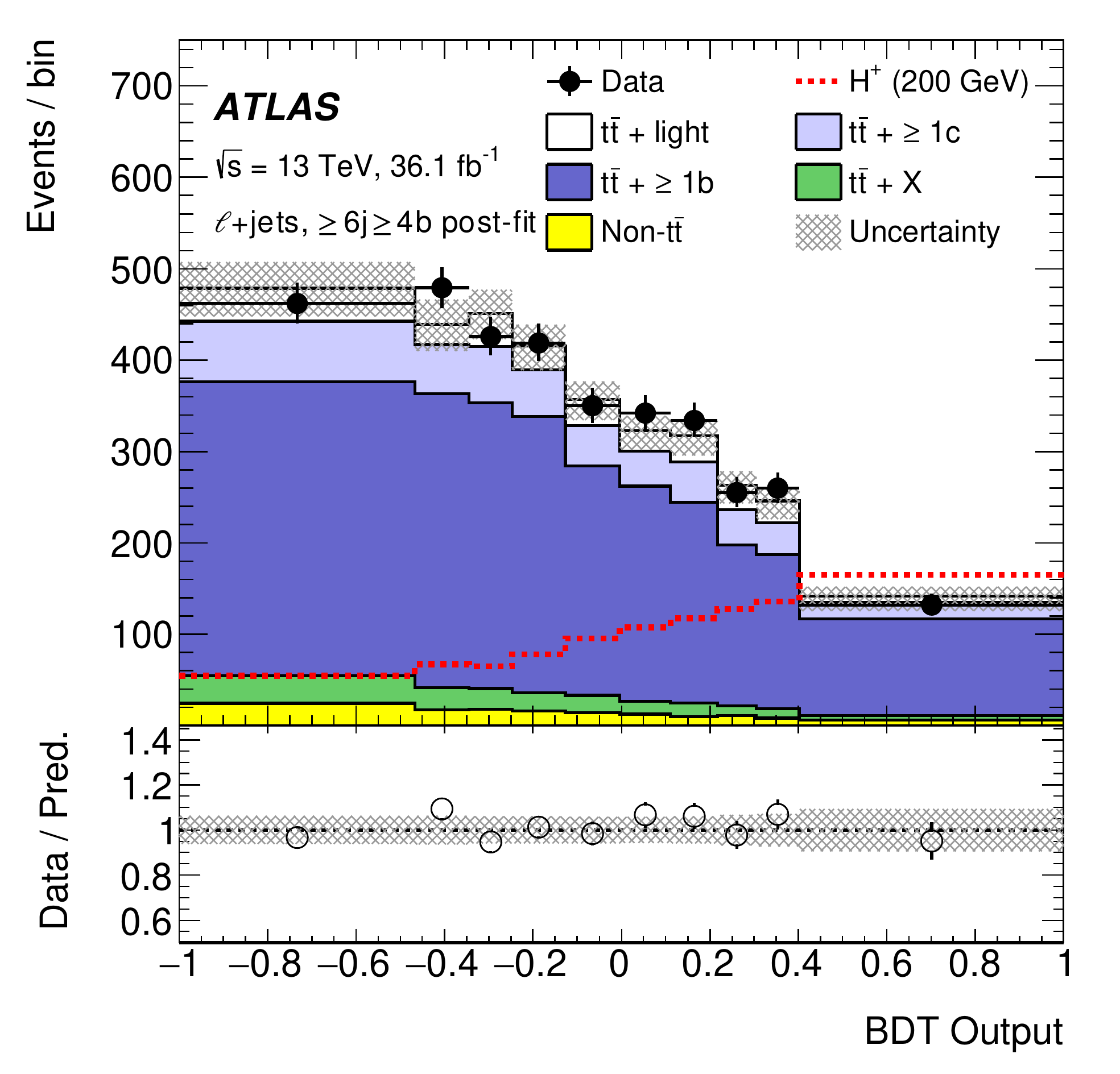} } \\
\end{center}
\caption {
Distributions of the BDT output after the fit to the data in the four SRs of the \ljets\ final state:
(a) \FiveJexThreeBex, 
(b) \SixJinThreeBex, 
(c) \FiveJexFourBin\ and 
(d) \SixJinFourBin\ 
for the 200~\GeV\ mass hypothesis.
Each background process is normalised according to its post-fit cross-section.
The $\ttbar + X$ includes contributions from \ttW, \ttZ\ and \ttH. 
The total prediction of the BDT distributions includes cases where the signal obtained from the fit is negative. 
For this particular mass point the fitted signal strength is $\mu = -0.4 \pm 1.5$~pb. 
The pre-fit signal distribution is shown superimposed as a dashed line with arbitrary normalisation.
The lower panels display the ratio of the data to the total prediction.
The hatched bands show the post-fit uncertainties.
\label{fig:BDTSingleLeptonPost200} }
\end{figure}

\begin{figure}[htb]
  \begin{center}
\subfloat[]{ \includegraphics[width=0.45\textwidth]{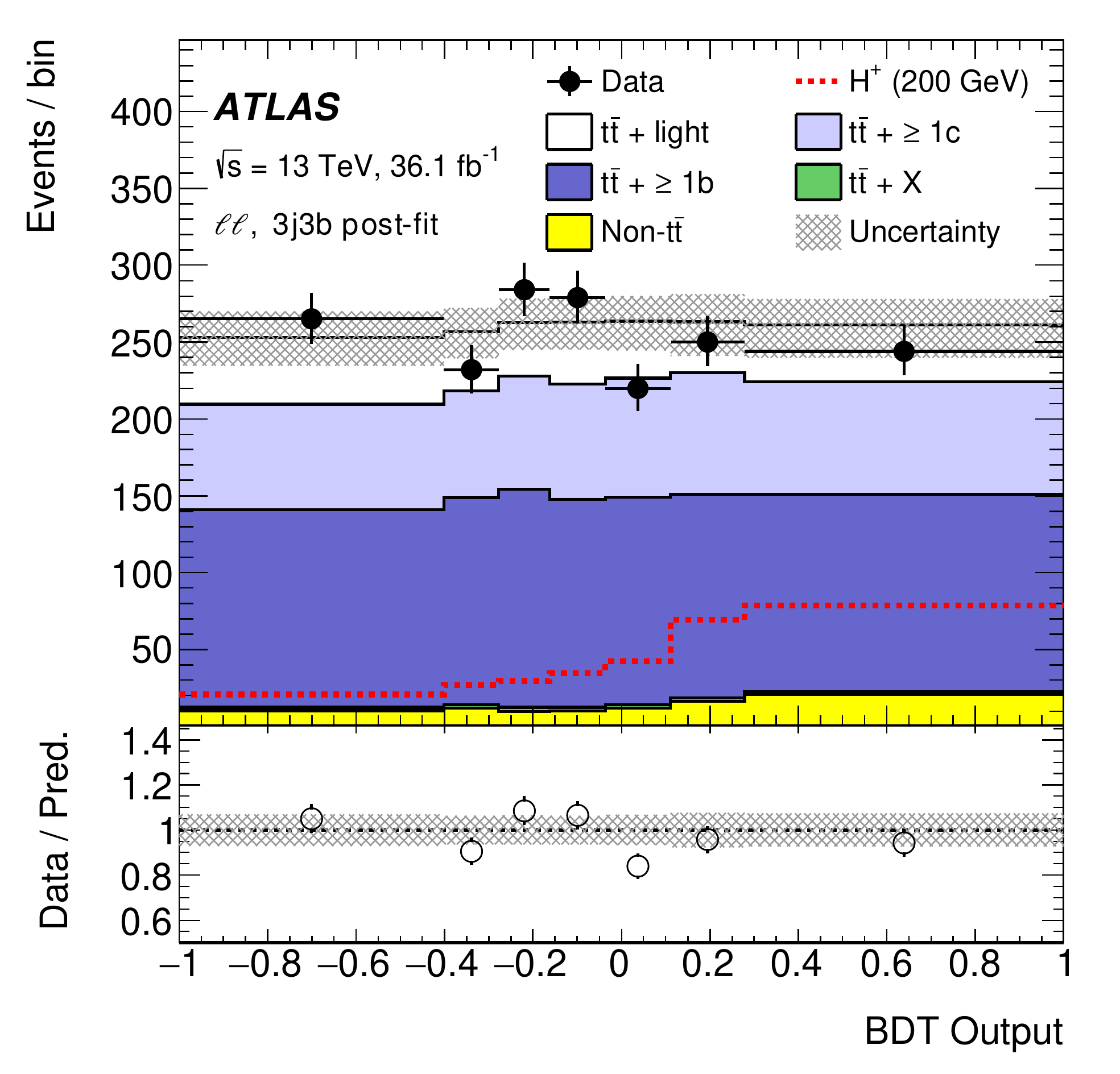} }
\subfloat[]{ \includegraphics[width=0.45\textwidth]{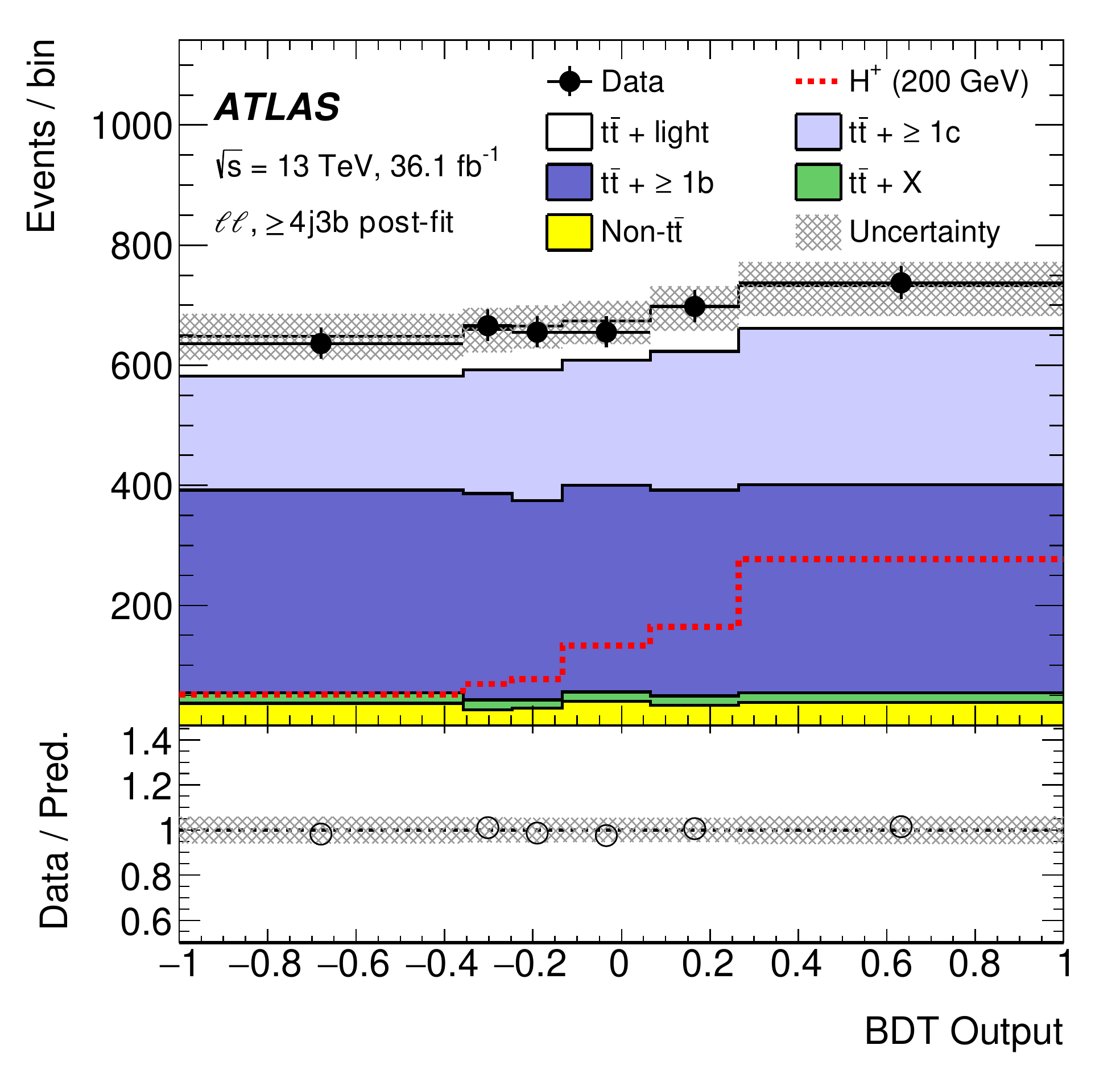} }\\
\subfloat[]{ \includegraphics[width=0.45\textwidth]{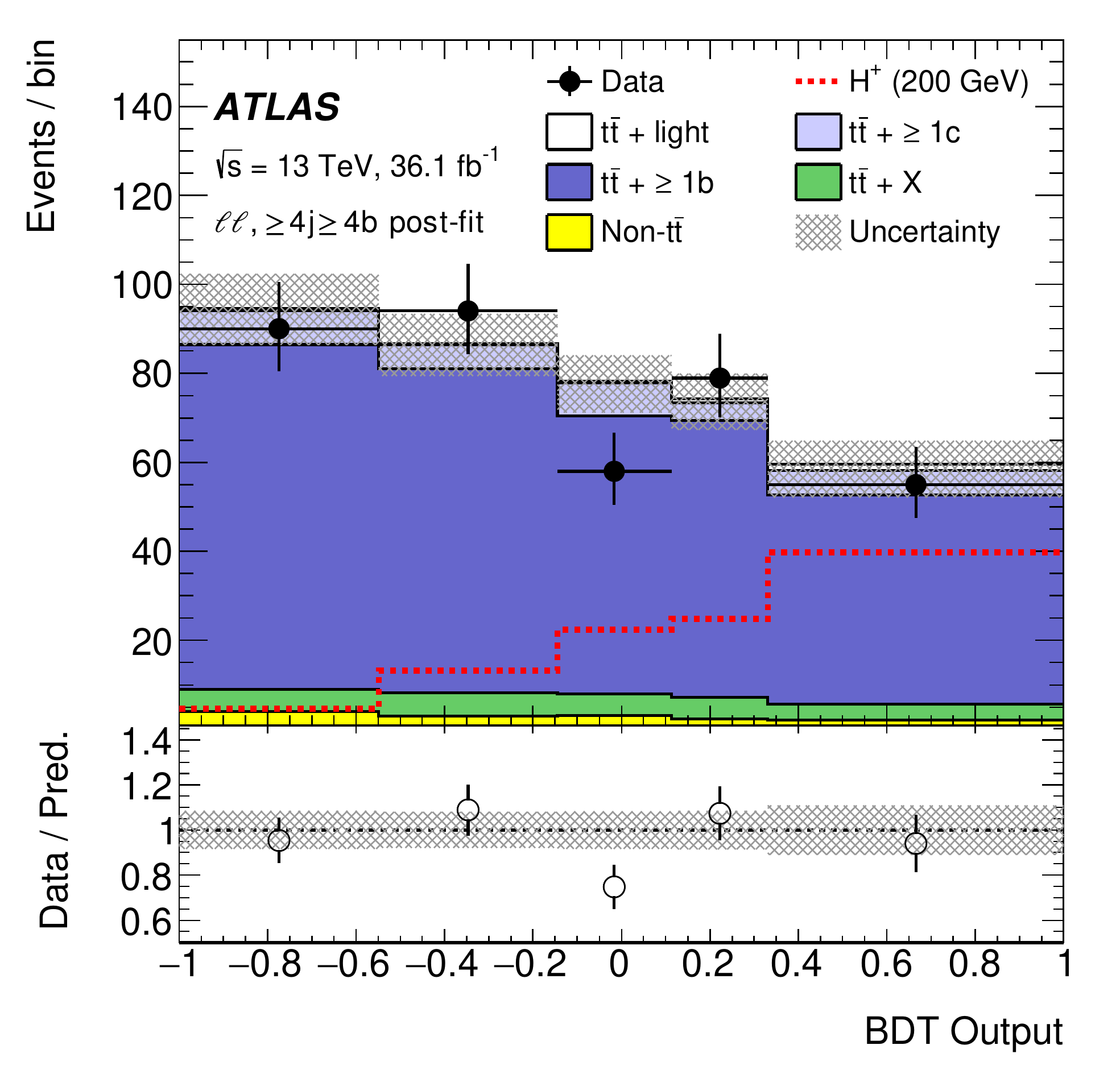} } \\
\end{center}
\caption {
Distributions of the BDT output after the fit to the data in the three SRs of the \dilepton\ final state:
(a) \ThreeJexThreeBex,
(b) \FourJinThreeBex\ and
(c) \FourJinFourBin\ 
for the 200~\GeV\ mass hypothesis.
Each background process is normalised according to its post-fit cross-section.
The $\ttbar + X$ includes contributions from \ttW, \ttZ\ and \ttH. 
The total prediction of the BDT distributions includes cases where the signal obtained from the fit is negative. 
For this particular mass point the fitted signal strength is $\mu = -0.4 \pm 1.5$~pb.
The pre-fit signal distribution is shown superimposed as a dashed line with arbitrary normalisation.
The lower panels display the ratio of the data to the total prediction.
The hatched bands show the post-fit uncertainties.
\label{fig:BDTDiLeptonPost200} }
\end{figure}

Tables~\ref{tab:postFitEventYieldsLJets} and \ref{tab:postFitEventYieldsDilepton2} show the post-fit event yields under the background-plus-signal hypothesis for a signal mass $\mhp = 200\,\GeV$. 
A value of $\sigma(pp \to  \tbhplus)\times \BR(\hplustb)=-0.36$~pb is obtained from the fit. 
The corresponding post-fit distributions of the BDT discriminant in the SRs are shown in Figures~\ref{fig:BDTSingleLeptonPost200} and~\ref{fig:BDTDiLeptonPost200} for a 200~\GeV\ \hplus\ mass hypotheses for the \ljets\ and \dilepton\ final state, respectively.

\begin{table}[htbp!]
\caption{Event yields of the SM background processes and data in all categories of the \ljets\ final state, after the fit to the data under the background-plus-signal hypothesis ($\mhp=200$~\GeV). 
The expected event yields for the \hplus\ signal masses of 200~\GeV\ and 800~\GeV\ are shown with pre-fit uncertainties and assuming a cross-section times branching ratio of 1~pb. The quoted uncertainties include both the statistical and systematic components. The uncertainties take into account correlations and constraints of the nuisance parameters. 
`Other top' includes contributions from $Zt$ as well as $s$- and $t$-channel single top production.\label{tab:postFitEventYieldsLJets}}
\footnotesize
\begin{center}
\begin{tabular}{l r@{\,$\pm$\,}l  r@{\,$\pm$\,}l  r@{\,$\pm$\,}l  r@{\,$\pm$\,}l  r@{\,$\pm$\,}l  r@{\,$\pm$\,}l }
Process  & \multicolumn{2}{c}{ CR \FiveJexTwoBex } & \multicolumn{2}{c}{ SR \FiveJexThreeBex } & \multicolumn{2}{c}{ SR \FiveJexFourBin } & \multicolumn{2}{c}{ CR \SixJinTwoBex } & \multicolumn{2}{c}{ SR \SixJinThreeBex } & \multicolumn{2}{c}{ SR \SixJinFourBin} \\
\hline
\hline
\ttgeb & 15 300 & 2300 & 7400 & 1000 & 750 & 110 & 17100 & 2800 & 11 100 & 1500 & 2410 & 260\\
\ttgec & 47 000 & 12 000 & 6400 & 1700 & 260 & 80 & 55 000 & 11 000 & 9400 & 2000 & 450 & 180\\
\ttbar\ + light & 226 000 & 11 000 & 12 200 & 1100 & 89 & 35 & 132 000 & 10 000 & 8500 & 1100 & 260 & 120\\
Non-prompt leptons & 15 000 & 6000 & 600 & 500 & 11 & 8 & 13 000 & 6000 & 700 & 400 & 4 & 5\\
\ttW & 340 & 50 & 29 & 4 & 0.66 & 0.22 & 540 & 80 & 72 & 11 & 5.0 & 1.2\\
\ttZ & 390 & 50 & 78 & 10 & 12.2 & 2.2 & 720 & 90 & 183 & 23 & 50 & 7\\
Single top $Wt$ & 8900 & 2400 & 690 & 210 & 23 & 13 & 5400 & 1800 & 640 & 260 & 53 & 31\\
Other top & 328 & 27 & 28.2 & 2.6 & 3.1 & 0.6 & 183 & 20 & 46 & 11 & 14 & 5\\
Diboson & 410 & 210 & 29 & 15 & 2.0 & 2.1 & 340 & 170 & 37 & 19 & 4.3 & 2.5\\
$W$ + jets & 9000 & 4000 & 540 & 240 & 16 & 9 & 5200 & 2100 & 470 & 200 & 27 & 12\\
$Z$ + jets & 2100 & 600 & 104 & 35 & 4.9 & 1.8 & 1300 & 400 & 130 & 40 & 11 & 4\\
\ttH & 252 & 24 & 127 & 13 & 30 & 4 & 520 & 50 & 315 & 32 & 117 & 16\\
\tH & 19.5 & 2.4 & 10.6 & 1.3 & 2.21 & 0.32 & 27.2 & 3.5 & 15.7 & 2.0 & 5.0 & 0.7\\
\hline
Total & 328 000 & 7000 & 28 400 & 900 & 1220 & 60 & 233 000 & 6000 & 31 800 & 800 & 3410 & 150\\
\hline
Data  & \multicolumn{1}{r}{ 334 813\,\,} && \multicolumn{1}{r}{ 29 322\,\,} && \multicolumn{1}{r}{ 1210\,\,} && \multicolumn{1}{r}{ 234 053\,\,} && \multicolumn{1}{r}{ 32 151\,\,} && \multicolumn{1}{r}{ 3459\,\,} & \\
\hline
\hplus\ (200~\GeV) & 470 & 50 & 220 & 23 & 25.3 & 3.3 & 340 & 50 & 235 & 34 & 60 & 9\\
\hplus\ (800~\GeV) & 630 & 90 & 390 & 70 & 56 & 12 & 1230 & 190 & 1020 & 170 & 350 & 70\\
\end{tabular}
\end{center}
\end{table}

\begin{table}[htbp!]
\caption{Event yields of the SM background processes and data in all categories of the \dilepton\ final state, after the fit to the data under the background-plus-signal hypothesis ($\mhp=200$~\GeV). 
The expected event yields for the \hplus\ signal masses of 200~\GeV\ and 800~\GeV\ are shown with pre-fit uncertainties and assuming a cross-section times branching ratio of 1~pb. The quoted uncertainties include both the statistical and systematic components. The uncertainties take into account correlations and constraints of the nuisance parameters. 
`Other top' includes contributions from $Zt$ as well as $s$- and $t$-channel single top production.\label{tab:postFitEventYieldsDilepton2}}
\footnotesize
\begin{center}
\begin{tabular}{l r@{\,$\pm$\,}l  r@{\,$\pm$\,}l  r@{\,$\pm$\,}l  r@{\,$\pm$\,}l  r@{\,$\pm$\,}l }
Process  & \multicolumn{2}{c}{ CR \ThreeJexTwoBex } & \multicolumn{2}{c}{ SR/CR \ThreeJexThreeBex } & \multicolumn{2}{c}{ CR \FourJinTwoBex } & \multicolumn{2}{c}{ SR \FourJinThreeBex } & \multicolumn{2}{c}{ SR \FourJinFourBin } \\
\hline
\hline
\ttgeb & 2330 & 330 & 940 & 130 & 3300 & 500 & 2050 & 280 & 322 & 35\\
\ttgec & 6100 & 1300 & 520 & 140 & 9900 & 2000 & 1310 & 290 & 30 & 14\\
\ttbar\ + light & 50 700 & 2300 & 260 & 70 & 32 500 & 2100 & 420 & 120 & 4 & 5\\
Non-prompt leptons & 420 & 110 & 6.7 & 2.4 & 620 & 160 & 48 & 13 & 2.2 & 0.8\\
\ttW & 48 & 7 & 1.48 & 0.17 & 129 & 7 & 9.8 & 1.1 & 0.55 & 0.21\\
\ttZ & 43 & 5 & 5.8 & 1.1 & 174 & 10 & 32.9 & 2.0 & 7.0 & 1.3\\
Single top $Wt$ & 1700 & 500 & 40 & 12 & 1110 & 330 & 63 & 26 & 3.9 & 2.0\\
Other top & 3.9 & 0.5 & 0.12 & 0.05 & 21.8 & 3.5 & 5.8 & 2.2 & 2.0 & 0.9\\
Diboson & 36 & 4 & 1.2 & 0.4 & 46 & 6 & 3.1 & 0.9 & 0.48 & 0.28\\
$Z$ + jets & 1600 & 500 & 42 & 16 & 1300 & 400 & 82 & 29 & 5.3 & 2.0\\
\ttH & 26.2 & 1.3 & 8.5 & 0.5 & 116 & 6 & 52.2 & 3.5 & 16.0 & 1.9\\
\tH & 1.95 & 0.27 & 0.42 & 0.10 & 5.7 & 0.7 & 2.14 & 0.32 & 0.48 & 0.09\\
\hline
Total & 62 800 & 2800 & 1810 & 110 & 49 300 & 2300 & 4060 & 200 & 390 & 28\\
\hline
Data  & \multicolumn{1}{r}{ 62 399\,\,} && \multicolumn{1}{r}{ 1774\,\,} && \multicolumn{1}{r}{ 48 356\,\,} && \multicolumn{1}{r}{ 4047\,\,} && \multicolumn{1}{r}{ 376\,\,} & \\
\hline
\hplus\ (200~\GeV) & 92 & 12 & 27 & 4 & 72 & 12 & 49 & 8 & 9.0 & 1.6\\
\hplus\ (800~\GeV) & 70 & 12 & 32 & 7 & 212 & 33 & 157 & 27 & 44 & 9\\
\end{tabular}
\end{center}
\end{table}

A summary of the systematic uncertainties is given in Table~\ref{tab:systematics}. 
Depending on the particular \hplus\ mass hypothesis, the total systematic uncertainty is dominated by the uncertainties in the modelling of the \ttbin\ background, the jet flavour-tagging uncertainties and the uncertainties due to the limited size of simulated event samples.

\begin{table}[htbp!]
  \caption{
    The summary of the effects of the systematic uncertainties on the signal strength parameter, $\mu=\sigma(pp \rightarrow \tbhplus) \times \BR(\hplustb)$, for the combination of the \ljets\ and \dilepton\ final states is shown for an \hplus\ signal with a mass of 200 and 800~\GeV. Due to correlations between the different sources of uncertainty, the total systematic uncertainty can be different from the sum in quadrature of the individual sources. The normalisation factors for both \ttgeb\ and \ttgec\ are included in the statistical component.
    The total uncertainty corresponds to a best-fit value of $\mu$ of $-0.4$~pb at $\mhp=200$~\GeV\ and $-0.02$~pb at $\mhp=800$~\GeV.
    The expected upper limit on $\mu$ is 3.05~pb at $\mhp=200$~\GeV\ and 0.26~pb at $\mhp=800$~\GeV. 
\label{tab:systematics}}
\begin{center}
\begin{tabular}{l | c@{\hspace{1cm}}r@{.}l@{\hspace{1cm}}c | c@{\hspace{1cm}}r@{.}l@{\hspace{1cm}}c }
  Uncertainty Source & \multicolumn{4}{c|}{$\Delta\mu(\hplus_{200})$ [pb]} & \multicolumn{4}{c}{$\Delta\mu(\hplus_{800})$ [pb]} \\
  \hline \hline
Jet flavour tagging                  & &0&70&    & &0&050& \\
\ttgeb modelling                     & &0&65&    & &0&008& \\
Jet energy scale and resolution      & &0&44&    & &0&031& \\
$\ttbar$+light modelling             & &0&44&    & &0&019& \\
MC statistics                        & &0&37&    & &0&044& \\
\ttgec modelling                     & &0&36&    & &0&032& \\
Other background modelling           & &0&36&    & &0&039& \\
Luminosity                           & &0&24&    & &0&010& \\
Jet-vertex assoc., pile-up modelling & &0&10&    & &0&006& \\
Lepton, \MET, ID, isol., trigger     & &0&08&    & &0&003& \\
\Hp\ modelling                       & &0&03&    & &0&006& \\
\hline
Total systematic uncertainty         & &1&4 &    & &0&11 & \\
\hline
\ttgeb normalisation                 & &0&61&    & &0&022& \\
\ttgec normalisation                 & &0&28&    & &0&012& \\
\hline
Total statistical uncertainty        & &0&69&    & &0&050& \\
\hline \hline
Total uncertainty                    & &1&5 &    & &0&12 & \\
\end{tabular}
\end{center}
\end{table}

\begin{figure}[h!]
  \begin{center}
     \includegraphics[width=0.64\textwidth]{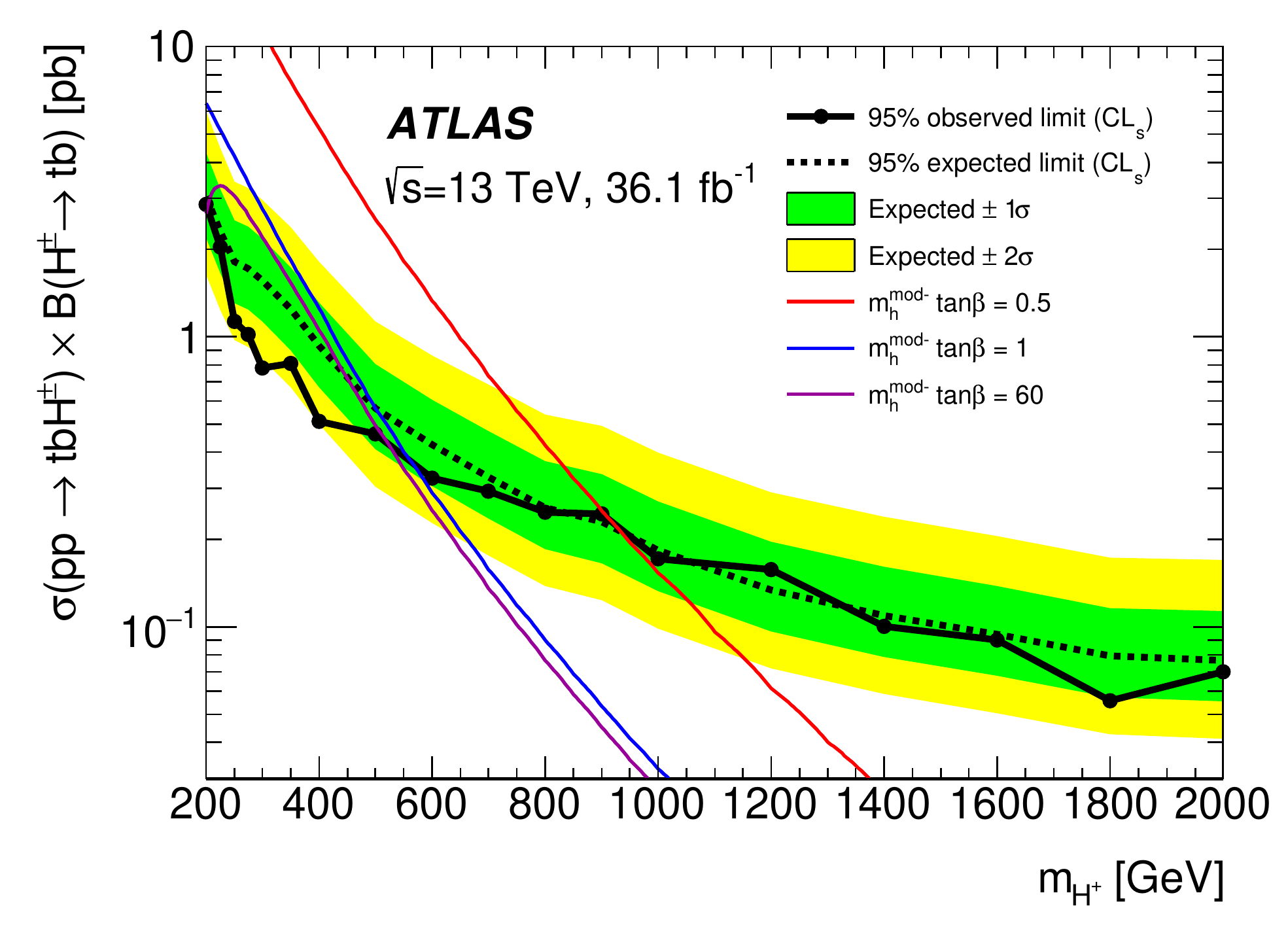}
  \end{center}
  \caption{
Expected and observed limits for the production of \hplustb\ in association with a top quark and a bottom quark. The bands surrounding the expected limit show the 68\% and 95\% confidence intervals. The limits are based on the combination of the \ljets\ and \dilepton\ final states.
Theory predictions are shown for three representative values of \tanbeta\ in the \mhmodm\ benchmark scenario~\cite{Carena:2013ytb}.
Uncertainties in the predicted \hplus\ cross-sections or branching ratios are not considered.
\label{fig:resultLimits}
}
\end{figure}

The 95\% confidence level (CL) upper limits on $\sigma(pp \to \tbhplus) \times \BR(\hplustb)$ using the $\mathrm{CL}_\mathrm{s}$ method are presented in Figure~\ref{fig:resultLimits}.
The observed (expected) 95\% CL upper limits on the $pp \to tb\hplus$ production cross-section times the branching ratio $\BR(\hplustb)$ range from $\sigma \times \BR = 2.9 \, (3.0)$~pb at $\mhp = 200\,\GeV$ to $\sigma \times \BR = 0.070 \, (0.077)$~pb at $\mhp = 2\,\TeV$. 
The compatibility of the SM hypothesis with the results obtained from the fit to the data is tested. 
The largest deviation from the SM hypothesis is observed at 300~\GeV. 
Given that a negative $\hat\mu$ is observed under this mass hypothesis, the test statistic $t_0  = -2 \ln {( \mathcal{L}( 0, \hat{\hat\theta}(0) ) / \mathcal{L}( \hat\mu, \hat\theta ) )}$ is used to quantify the deviation of the fitted result from the SM expectation. 
A local $p_0$ value of 1.13\% is obtained at 300~\GeV, corresponding to the probability to obtain a deviation at least as large as the one observed in data provided that only SM processes are present.

\begin{figure}[h!]
  \begin{center}
    \subfloat[]{ \includegraphics[width=0.49\textwidth]{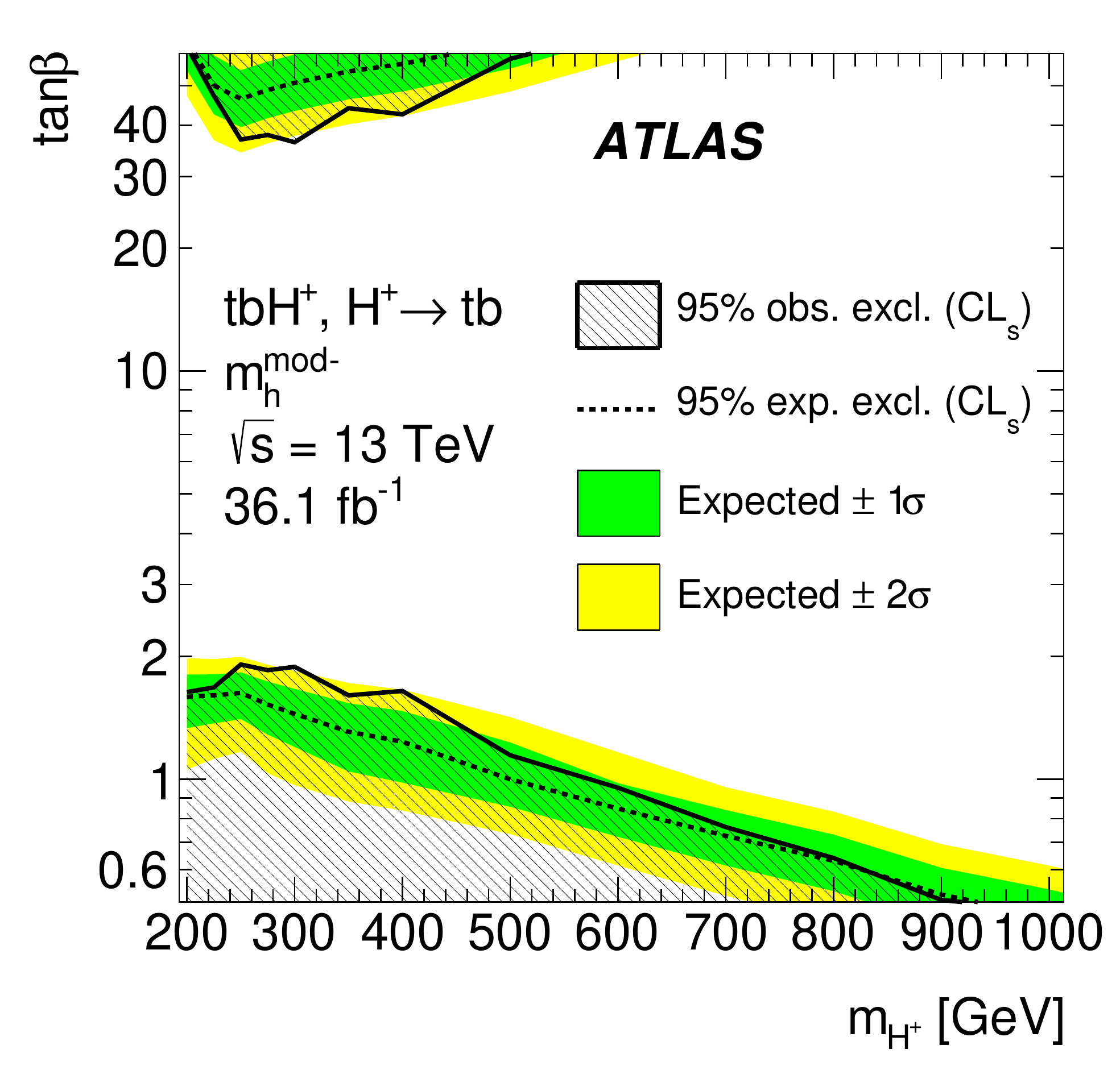}}
    \subfloat[]{ \includegraphics[width=0.49\textwidth]{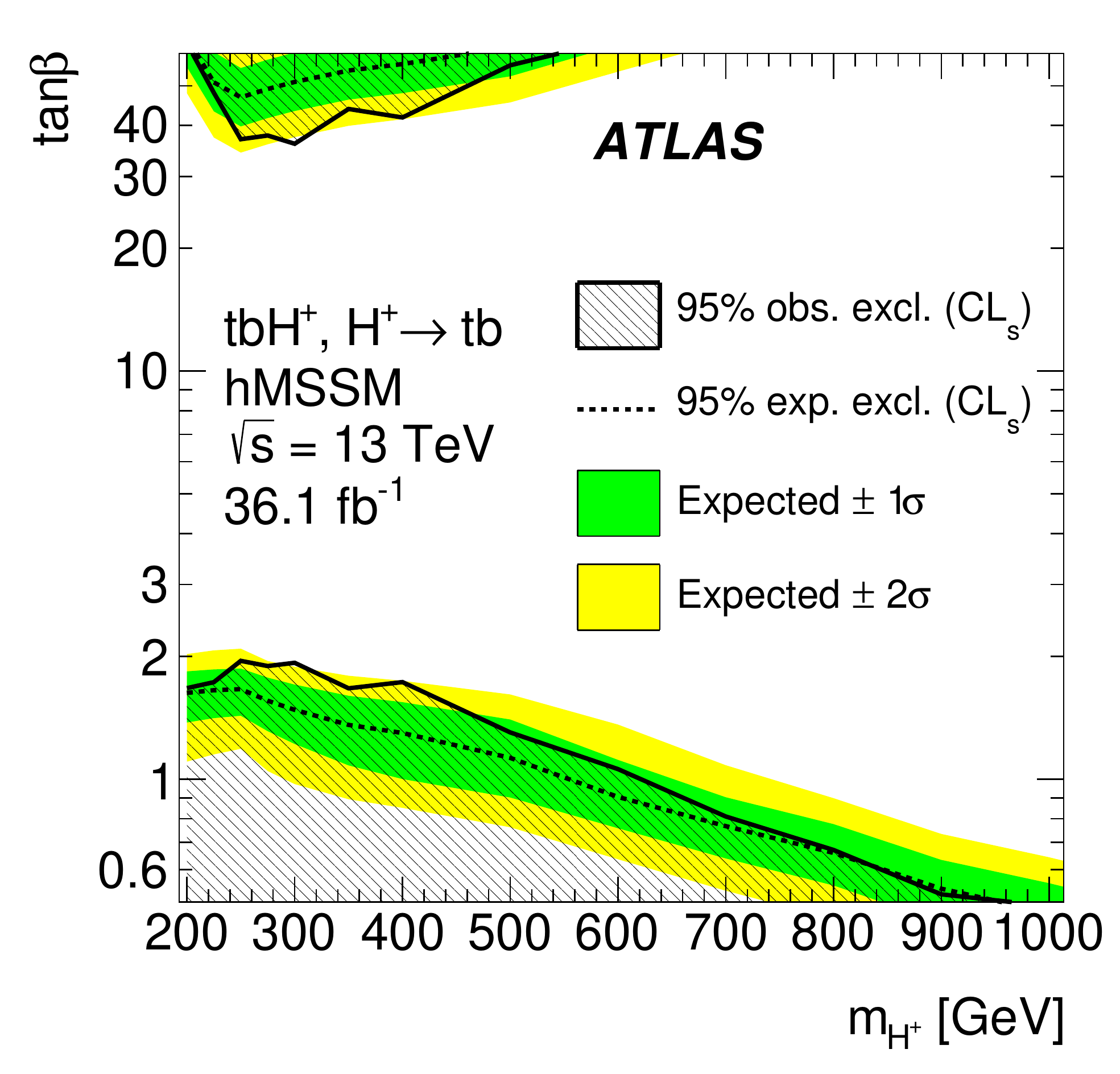}}
  \end{center}
  \caption{
Expected and observed limits on \tanbeta\ as a function of $\mhp$ in the \mhmodm~\cite{Carena:2013ytb} (left) and the hMSSM~\cite{Djouadi:2013uqa} (right) scenarios of the MSSM.
Limits are shown for \tanbeta\ values in the range of 0.5--60, where predictions are available from both scenarios.
The bands surrounding the expected limits show the 68\% and 95\% confidence intervals.
The limits are based on the combination of the \ljets\ and \dilepton\ final states. 
The production cross-section of \ttH\ and \tH, as well as the branching ratios of the \Higgs, are fixed to their SM values at each point in the plane. 
Uncertainties in the predicted \hplus\ cross-sections or branching ratios are not considered.
\label{fig:hMSSM_mhmodm}}
\end{figure}

Figure~\ref{fig:hMSSM_mhmodm} shows 95\% CL exclusion limits set on \tanbeta\ for the \mhmodm\ scenario of the MSSM~\cite{Carena:2013ytb,Heinemeyer:2013tqa,deFlorian:2016spz} 
and the hMSSM~\cite{Djouadi:2013uqa,Djouadi:2015jea,Bagnaschi:2015hka}. 
Beyond tree level, the Higgs sector is affected by the choice of parameters in addition to Higgs boson masses and \tanbeta. For the  \mhmodm\ benchmark scenario the top-squark mixing parameter is chosen such that the mass of the lightest CP-even Higgs boson, $m_h$, is close to the measured mass of the Higgs boson that was discovered at the LHC\@.
In the hMSSM scenario, instead of adjusting the parameters of soft supersymmetry breaking, the value of $m_h$ is used to predict the masses and couplings of the MSSM Higgs bosons.

For \hplus\ masses of 200--920~\GeV\ (200--965~\GeV), the observed exclusion of low values of \tanbeta\ at 95\% CL is in the range 0.5--1.91 (0.5--1.95) for the \mhmodm\ (hMSSM) scenario. 
The most stringent limits on \tanbeta\ are set for \hplus\ masses around 250~\GeV. 
High values of \tanbeta\ between 36 and 60 are excluded in the \hplus\ mass range 200--520~\GeV\ (220--540~\GeV) for the \mhmodm\ (hMSSM) scenario. 
The most stringent exclusion, $\tanbeta>36$, is at 300~\GeV\ for both the \mhmodm\ and hMSSM benchmark scenarios.
In the \mhmodm\ scenario for $\tanbeta = 0.5$, the observed (expected) exclusion of \hplus\ masses is $\mhp < 920$~\GeV\ ($\mhp < 930$~\GeV). 

In comparison with a previous search for $t[b]\hplus$ production followed by $\hplus \rightarrow tb$ decays~\cite{Aad:2015typ},   
more stringent limits on \hplus\ masses for particular models and parameter choices can be set. 
The analysis reach is increased and now also includes \Hplus\ masses between 600~\GeV\ and 2~\TeV. 
The excluded region of parameter space for the model-dependent interpretation is extended significantly for low \tanbeta\ and an additional excluded region is added at high \tanbeta.

\begin{figure}[h!]
  \begin{center}
    \subfloat[]{ \includegraphics[width=0.49\textwidth]{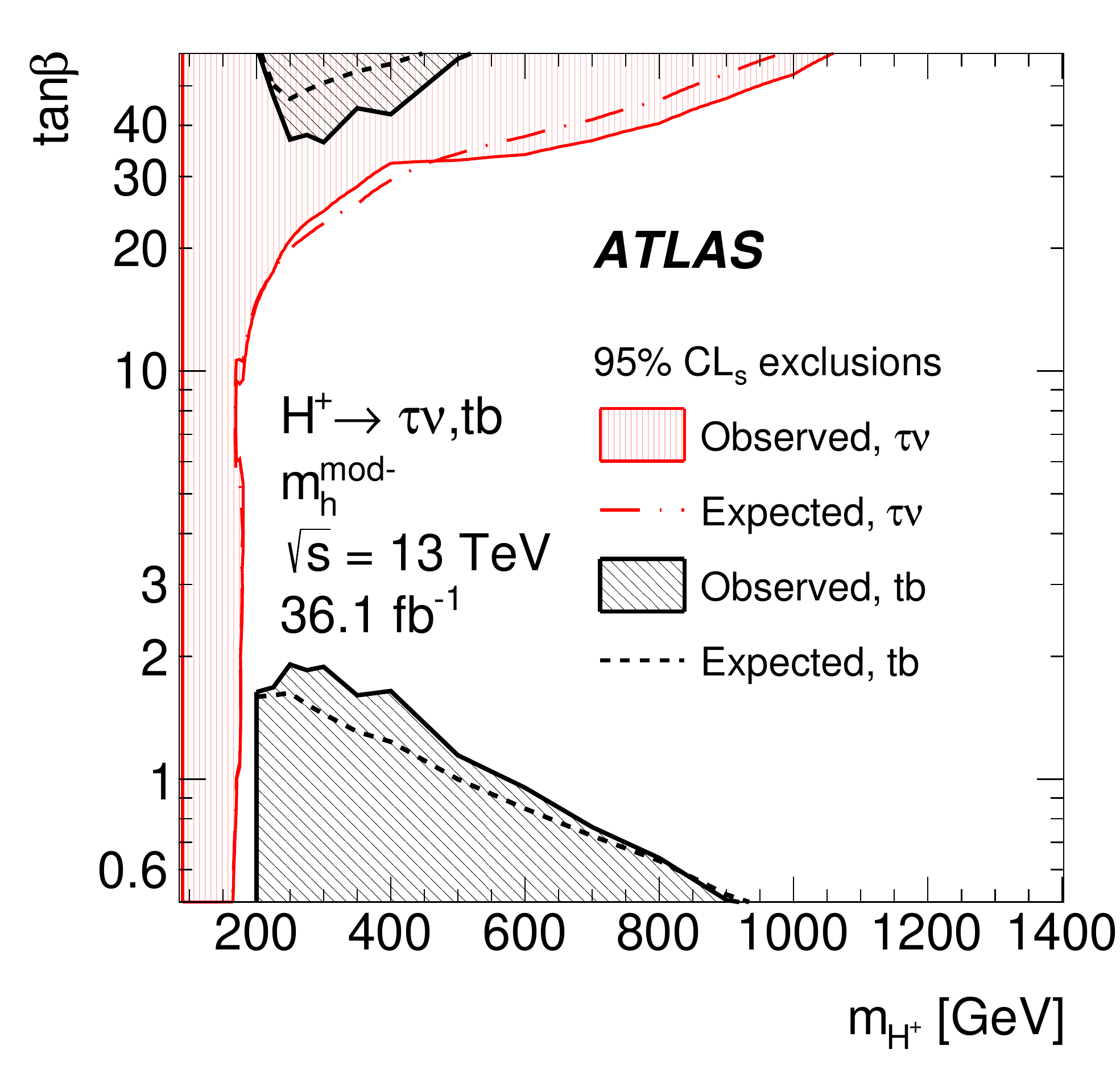}}
    \subfloat[]{ \includegraphics[width=0.49\textwidth]{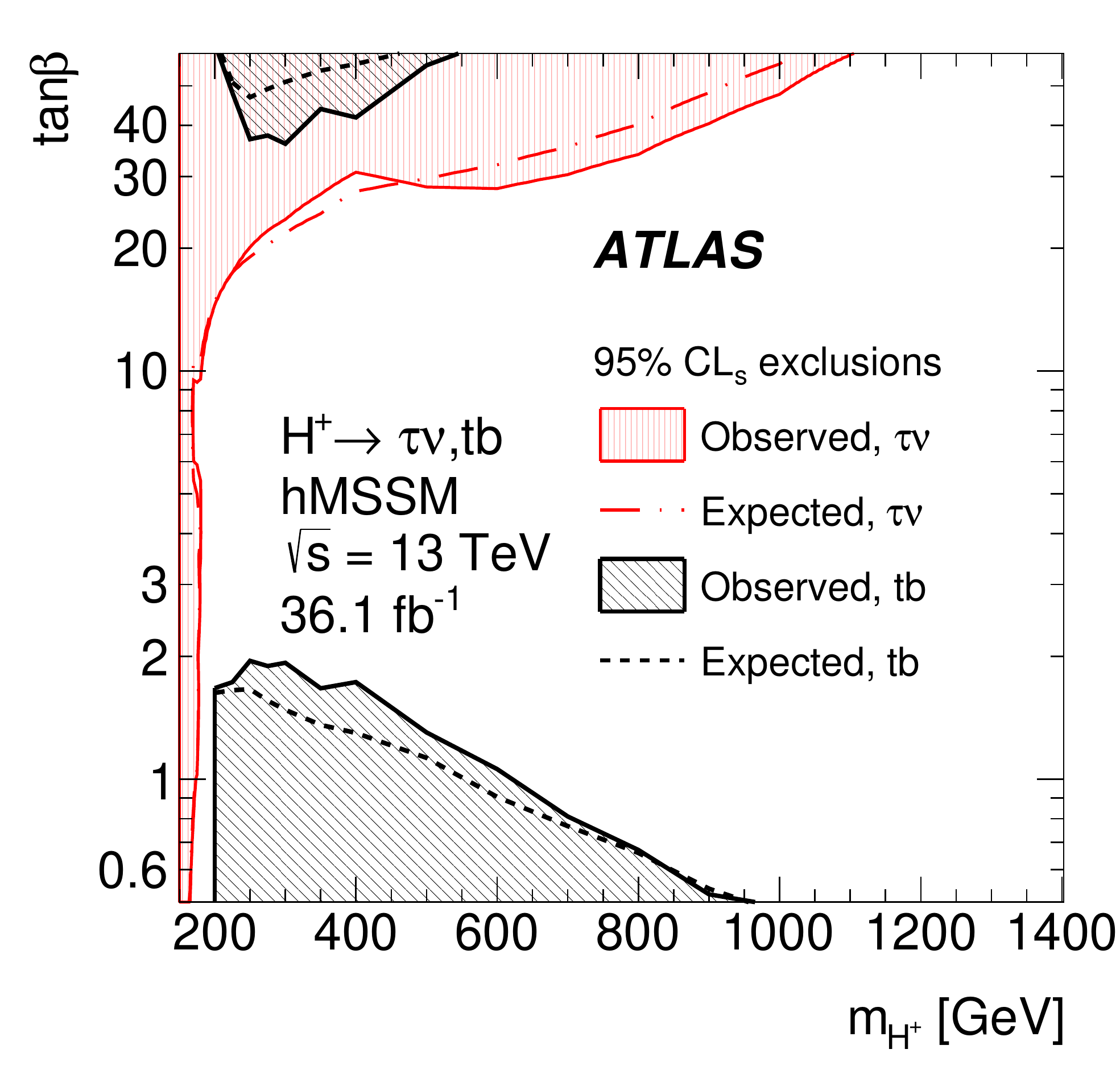}}
  \end{center}
  \caption{Expected and observed limits on \tanbeta\ as a function of $\mhp$ in the \mhmodm~\cite{Carena:2013ytb} (left) and the hMSSM~\cite{Djouadi:2013uqa} (right) scenarios of the MSSM.
Limits are shown for \tanbeta\ values in the range of 0.5--60, where predictions are available from both scenarios.
The limits are a superposition of the results obtained in the analysis presented here, and the ATLAS limits derived from the $\hplus \to \tau \nu$ decay~\cite{ATLAS-HIGG-2016-11}.
The expected limits from the $\tau \nu$ final state are shown as the horizontally hatched area, with the observed limit as a dash-dotted curve.
The expected limits from the $tb$ final state are shown as diagonally hatched areas, with the observed limit as dashed lines. 
At low \tanbeta, the strongest limits are from the $tb$ final state, whereas the exclusions at high \tanbeta\ and low \hplus\ masses are obtained from the $\tau \nu$ final state.
The exclusion limits for the hMSSM scenario are shown only for $\mhp > 150$~\GeV, where the corresponding theoretical predictions are available.
\label{fig:superposition_limits}}
\end{figure}

The ATLAS Collaboration has also set limits on the production of \hplus\ using the $\hplus \to \tau \nu$ decay with the same data~\cite{ATLAS-HIGG-2016-11}. 
The $\tau \nu$ final state can be used to set limits at high \tanbeta\ which are more stringent than those from the $tb$ final state, 
and to probe \hplus\ masses below 200~\GeV, in both the \mhmodm\ and hMSSM scenarios. 
Figure~\ref{fig:superposition_limits} shows a superposition of the limits from the two final states, 
where the limits from the $\tau\nu$ final state exclude a larger portion of the parameter space at high \tanbeta\ and low \hplus\ masses than the $tb$ limits alone. 

 \FloatBarrier

%% file: conclusions.tex
\section{Conclusions}
\label{sec:conclusion}
A search for charged Higgs bosons is performed using a data sample
corresponding to an integrated luminosity of \lumi\ from $pp$ collisions at $\sqrt{s} = 13\,\TeV$,
recorded by the ATLAS detector at the LHC\@.
The search for $pp\to tb\hplus$ is performed in the \hplus\ mass range 200--2000~\GeV.
The analysis uses multivariate techniques in the signal regions to enhance the separation of signal from background and utilises control regions to reduce the effect of large uncertainties in background predictions.

No significant excess above the expected SM background is found and observed (expected) 95\% CL upper limits are set on the $pp \to tb\hplus$ production cross-section times the branching ratio $\BR(\hplustb)$, which range from $\sigma \times \BR$ = 2.9  (3.0)~pb at $\mhp = 200\,\GeV$ to $\sigma \times \BR$ = 0.070 (0.077)~pb at $\mhp = 2\,\TeV$.

In the context of the \mhmodm\ (hMSSM) scenario of the MSSM, some values of \tanbeta, in the range 0.5--1.91 (0.5--1.95), are excluded for \hplus\ masses of 200--920 (200--965)~\GeV. 
For \hplus\ masses between 200 and 520~\GeV\ (220 and 540~\GeV), high values of \tanbeta\ are excluded, e.g.~$\tanbeta > 36$ is excluded at 300~\GeV.

Additionally, taking into consideration the $\hplus \to \tau \nu$ decay, even stricter exclusions can be made at high \tanbeta\ and low \hplus\ masses. 
In the context of the hMSSM, the \hplus\ mass range up to 1100 \GeV\ is excluded at $\tanbeta = 60$, and all \tanbeta\ values are excluded for \mhp\ below 160 \GeV.

%% file: acknowledgements/Acknowledgements.tex

We thank CERN for the very successful operation of the LHC, as well as the
support staff from our institutions without whom ATLAS could not be
operated efficiently.

We acknowledge the support of ANPCyT, Argentina; YerPhI, Armenia; ARC, Australia; BMWFW and FWF, Austria; ANAS, Azerbaijan; SSTC, Belarus; CNPq and FAPESP, Brazil; NSERC, NRC and CFI, Canada; CERN; CONICYT, Chile; CAS, MOST and NSFC, China; COLCIENCIAS, Colombia; MSMT CR, MPO CR and VSC CR, Czech Republic; DNRF and DNSRC, Denmark; IN2P3-CNRS, CEA-DRF/IRFU, France; SRNSFG, Georgia; BMBF, HGF, and MPG, Germany; GSRT, Greece; RGC, Hong Kong SAR, China; ISF and Benoziyo Center, Israel; INFN, Italy; MEXT and JSPS, Japan; CNRST, Morocco; NWO, Netherlands; RCN, Norway; MNiSW and NCN, Poland; FCT, Portugal; MNE/IFA, Romania; MES of Russia and NRC KI, Russian Federation; JINR; MESTD, Serbia; MSSR, Slovakia; ARRS and MIZ\v{S}, Slovenia; DST/NRF, South Africa; MINECO, Spain; SRC and Wallenberg Foundation, Sweden; SERI, SNSF and Cantons of Bern and Geneva, Switzerland; MOST, Taiwan; TAEK, Turkey; STFC, United Kingdom; DOE and NSF, United States of America. In addition, individual groups and members have received support from BCKDF, CANARIE, CRC and Compute Canada, Canada; COST, ERC, ERDF, Horizon 2020, and Marie Sk{\l}odowska-Curie Actions, European Union; Investissements d' Avenir Labex and Idex, ANR, France; DFG and AvH Foundation, Germany; Herakleitos, Thales and Aristeia programmes co-financed by EU-ESF and the Greek NSRF, Greece; BSF-NSF and GIF, Israel; CERCA Programme Generalitat de Catalunya, Spain; The Royal Society and Leverhulme Trust, United Kingdom. 

The crucial computing support from all WLCG partners is acknowledged gratefully, in particular from CERN, the ATLAS Tier-1 facilities at TRIUMF (Canada), NDGF (Denmark, Norway, Sweden), CC-IN2P3 (France), KIT/GridKA (Germany), INFN-CNAF (Italy), NL-T1 (Netherlands), PIC (Spain), ASGC (Taiwan), RAL (UK) and BNL (USA), the Tier-2 facilities worldwide and large non-WLCG resource providers. Major contributors of computing resources are listed in Ref.~\cite{ATL-GEN-PUB-2016-002}.

%% file: appendix.tex
\section{BDT input variables}
\label{sec:appendix}

In this appendix, the full list of variables used as inputs to the BDTs, described in Section~\ref{sec:analysis}, is reported. 

\begin{table}[h!]
\caption{Input variables to the classification BDT in the \ljets\ and \dilepton\ channels. 
The symbols $j$, $b$, $u$, $\ell$ and \met\ represent the four-momenta of jets, $b$-tagged jets, non-$b$-tagged jets, the lepton and the missing transverse momentum. 
All numbered indices refer to ordering in transverse momentum, with 1 as leading. 
The SRs where the variables are used are indicated for the \dilepton\ channel. 
In the \ljets\ channels the variables are used for all channels. In the \ljets\ channel the discriminant, $D$, is only used for charged Higgs boson masses $\mhp \leq 300\,\GeV$. For the \dilepton\ channel a very large set of kinematic variables using combinations of the analysis objects was examined, and approximately ten optimal variables were selected for each SR independently for the low-mass region ($\mhp \leq 600$~\GeV) and the high-mass region ($\mhp > 600$~\GeV). \label{tab:BDTvariables}}
\tiny
\begin{center}
  \begin{tabular}{l l c c c}
\ljets\ channel\\
\hline
\hline
$\pt (j_1)$                             &  Leading jet transverse momentum          \\
$m(b\text{-pair}^{\Delta R^{\text{min}}})$          &  Invariant mass of pair of $b$-tagged jets with smallest $\Delta R$ \\
$\pt(j_5)$                             &  Transverse momentum of fifth jet         \\ 
$H_2$                                        &  Second Fox--Wolfram moment \cite{Bernaciak2013} calculated using all jets and leptons \\
$\Delta R^{\text{avg}}(b\text{-pair})$         &  Average $\Delta R$ between all $b$-tagged jet pairs in the event \\
$\Delta R(\ell,b\text{-pair}^{\Delta R^{\text{min}}})$ &  $\Delta R$ between the lepton and the $b$-tagged jet pair with smallest $\Delta R$ \\
$m(u\text{-pair}^{\Delta R^{\text{min}}})$          &  Invariant mass of the non-$b$-tagged jet-pair with minimum $\Delta R$ \\
\Htjets                                      &  Scalar sum of all jets transverse momenta \\
$m(b\text{-pair}^{\pt^{\text{max}}})$         &  Invariant mass of the $b$-tagged jet pair with maximum transverse momentum  \\
$m^{\text{max}}(b\text{-pair})$                &  Largest invariant mass of any two $b$-tagged jets  \\
$m^{\text{max}}(j\text{-triplet})$             &  Largest invariant mass of any three jets  \\
$D$                                          &  Kinematic discriminant based on mass templates (for $\mhp \leq 300\,\GeV$) \\
\\
\dilepton\ channel, $m\leq 600\,\GeV$ & & \ThreeJexThreeBex  & \FourJinThreeBex  & \FourJinFourBin \\
\hline
\hline
$m((j,b)^{\pt^{\text{max}}})$                       &  Inv.~mass of the jet and $b$-tagged jet with largest \pt\         &  \checkmark  &              &  \\
$\Delta E(j_3,\ell_2)$                                &  Energy difference between the third jet and the subleading lepton               &  \checkmark  &              &  \\
$E(j_3)$                                           &  Energy of third jet                                                             &  \checkmark  &              &  \\
$\Delta m(j_1+j_2,j_1+j_3+\ell_2+\met)$  &  Inv.~mass difference between $j_1+j_2$ and $j_1+j_3+\ell_2+\met$  &  \checkmark  &              &  \\
$\Delta R(j_2,j_1+\ell_2+\met)$       &  Angular difference between subleading jet and $j_1+\ell_2+\met$  &  \checkmark  &      &  \\
$\pt(b_1)$                                   &  \pt\ of leading $b$-tagged jet                                           &  \checkmark  &             &  \\
$\pt((\ell,b)^{\Delta\eta^{\text{max}}})$              &  \pt\ of the pair of lepton and $b$-tagged jet with largest $\Delta \eta$        &  \checkmark  &             &  \\
$m((\ell,b)^{\Delta\phi^{\text{min}}})$                      &  Inv.~mass of the pair of lepton and $b$-tagged jet with smallest $\Delta \phi$            &              & \checkmark  &  \\
$\Delta E(b_1,\ell_1+\met)$           &  Energy difference between the leading $b$-tagged jet and $\ell_1+\met$        &              & \checkmark  &  \\
$\Delta m(j_2+j_3,j_1+\ell_1+\ell_2)$                      &  Inv.~mass difference between $j_2+j_3$ and $j_1+\ell_1+\ell_2$                     &              & \checkmark  &  \\
$\Delta m(\ell_1+j_3+\met,j_1+j_2+\ell_2)$ &  Inv.~mass difference between $\ell_1+j_3+\met$ and $j_1+j_2+\ell_2$   &              & \checkmark  &  \\
$\Delta \pt(j_1,j_3)$                        &  \pt\ difference between leading and third jet                     &              & \checkmark  &  \checkmark  \\
$m^{\text{min}}(b\text{-pair})$                      &  Smallest invariant mass of any $b$-tagged jet pair                                      &              & \checkmark  &  \checkmark  \\
$m^{\text{min}}(\ell,b)$                                &  Smallest invariant mass of any pair of lepton and $b$-tagged jet                               &              & \checkmark  &  \checkmark  \\
$\pt(b_2+\ell_1+\ell_2+\met)$     &  \pt\ of $b_2+\ell_1+\ell_2+\met$                           &              &             &  \checkmark  \\
$\Delta R(\ell_2,j_2+j_3+\ell_1+\met)$  &  Angular difference between $\ell_2$ and $j_2+j_3+\ell_1+\met$   &      &      &  \checkmark  \\
$H_{\text T}^{\text{all}}$                                 &  Scalar sum of all jets and leptons transverse energy          &              &             & \checkmark  \\
\\
\dilepton\ channel, $m> 600\,\GeV$            &                                            & \ThreeJexThreeBex  & \FourJinThreeBex  & \FourJinFourBin \\
\hline
\hline
$\pt((\ell,b)^{\Delta\eta^{\text{min}}})$           &  \pt\ of the pair of lepton and $b$-tagged jet with smallest $\Delta\eta$   & \checkmark &             & \checkmark  \\
$\Delta \pt(j_1,j_3)$                    &  \pt\ difference between leading and third jets           & \checkmark &             & \checkmark  \\ 
$\Delta m(j_2+\ell_1+\met,j_1+j_3+\ell_1)$  &  Inv.~mass difference between $j_2+\ell_1+\met$ and $j_1+j_3+\ell_1$   & \checkmark &    &  \\
$\pt((\ell,b)^{\Delta R^{\text{min}}})$           &  \pt\ of the pair of lepton and $b$-tagged jet with smallest $\Delta R$ & \checkmark &             &  \\
$m(j\text{-pair}^{\Delta \eta^{\text{min}}})$           &  Inv.~mass of the jet pair with smallest $\Delta \eta$              & \checkmark &             &  \\
$\Delta \pt (j_1,j_2+\met)$  &  \pt\ difference between leading jet and $j_2+\met$   & \checkmark  &  &  \\
$\pt(j_1+j_2+j_3+\ell_1)$                   &  \pt\  of $j_1+j_2+j_3+\ell_1$                                & \checkmark &             &  \\
$\Delta E(\ell_1+\met,j_1+j_2)$   &  Energy difference between $\ell_1+\met$ and $j_1+j_2$         & \checkmark &             &  \\
$E(j_1)$                                       &  Energy of the leading jet                                               & \checkmark & \checkmark  &  \\
$\pt^{\text{max}}(j\text{-pair})$          &  Maximum \pt\ of any jet pair                             & \checkmark & \checkmark  &  \\
$m(b_1+b_2+\ell_1+\ell_2+\met)$      &  Inv.~mass of $b_1+b_2+\ell_1+\ell_2+\met$                   &            & \checkmark  &  \\
$\pt((\ell,b)^{\Delta\eta^{\text{min}}})$          &  \pt\ of the lepton-$b$-jet pair with smallest separation in $\eta$ &  & \checkmark  &  \\    
$\Delta \pt(\ell_2,u_1+b_2+\met)$  &  \pt\ difference between subleading lepton and $u_1+b_2+\met$ & & \checkmark &  \\
$\Delta \pt(\ell_2,u_1+b_1+\met)$  &  \pt\ difference between subleading lepton and $u_1+b_1+\met$ & & \checkmark &  \\
$\Delta \pt(\ell_2,\ell_1+\met)$      &  \pt\ difference between subleading lepton and $\ell_1+\met$ & & \checkmark &  \\
$\Delta \pt(j_1,j_3+\ell_1+\met)$   &  \pt\ difference between leading jet and $j_3+\ell_1+\met$ & & \checkmark &  \\
$\Delta E(\ell_1,j_2+\met)$               &  Energy difference between leading lepton and $j_2+\met$    &           & \checkmark     &  \\
$m^{\text{min}}(b\text{-pair})$                  &  Smallest invariant mass of any $b$-tagged jet pair                             &            & \checkmark  & \checkmark  \\
\Htall                              &  Scalar sum of all jets and leptons transverse momenta &           &               & \checkmark  \\
$\pt(j_3+\ell_1)$   &  \pt\ of $j_3+\ell_1$                                         &           &              & \checkmark  \\ 
$\Delta \pt(b_2,b_1+\ell_2)$  &  \pt\ difference between subleading $b$-tagged jet and $b_1+\ell_2$  &           &              & \checkmark  \\
$\Delta \pt(j_2,j_3+\ell_1+\met)$   &  \pt\ difference between subleading jet and $j_3+\ell_1+\met$  & &          & \checkmark  \\
$\Delta E(j_3,j_2+\ell_1+\ell_2+\met)$       &  Energy difference between third jet and $j_2+\ell_1+\ell_2+\met$  &             &            & \checkmark  \\
$\Delta m(j_2+\ell_2+\met,j_1+\ell_2+\met)$  &  Inv.~mass difference between $j_2+\ell_2+\met$ and $j_1+\ell_2+\met$ &  &  & \checkmark  \\

\end{tabular}
\end{center}
\end{table}

%% file: atlas_authlist.tex
 
\begin{flushleft}
{\Large The ATLAS Collaboration}

\bigskip

M.~Aaboud$^\textrm{\scriptsize 34d}$,    
G.~Aad$^\textrm{\scriptsize 99}$,    
B.~Abbott$^\textrm{\scriptsize 124}$,    
O.~Abdinov$^\textrm{\scriptsize 13,*}$,    
B.~Abeloos$^\textrm{\scriptsize 128}$,    
D.K.~Abhayasinghe$^\textrm{\scriptsize 91}$,    
S.H.~Abidi$^\textrm{\scriptsize 164}$,    
O.S.~AbouZeid$^\textrm{\scriptsize 39}$,    
N.L.~Abraham$^\textrm{\scriptsize 153}$,    
H.~Abramowicz$^\textrm{\scriptsize 158}$
% ...
\bigskip
\\

% ...
$^{13}$Institute of Physics, Azerbaijan Academy of Sciences, Baku; Azerbaijan.\\
% ...
$^{34}$$^{(a)}$Facult\'e des Sciences Ain Chock, R\'eseau Universitaire de Physique des Hautes Energies - Universit\'e Hassan II, Casablanca;$^{(b)}$Centre National de l'Energie des Sciences Techniques Nucleaires (CNESTEN), Rabat;$^{(c)}$Facult\'e des Sciences Semlalia, Universit\'e Cadi Ayyad, LPHEA-Marrakech;$^{(d)}$Facult\'e des Sciences, Universit\'e Mohamed Premier and LPTPM, Oujda;$^{(e)}$Facult\'e des sciences, Universit\'e Mohammed V, Rabat; Morocco.\\
% ...
$^{39}$Niels Bohr Institute, University of Copenhagen, Copenhagen; Denmark.\\
% ...
$^{91}$Department of Physics, Royal Holloway University of London, Egham; United Kingdom.\\
% ...
$^{99}$CPPM, Aix-Marseille Universit\'e, CNRS/IN2P3, Marseille; France.\\
% ...
$^{124}$Homer L. Dodge Department of Physics and Astronomy, University of Oklahoma, Norman OK; United States of America.\\
% ...
$^{128}$LAL, Universit\'e Paris-Sud, CNRS/IN2P3, Universit\'e Paris-Saclay, Orsay; France.\\
% ...
$^{153}$Department of Physics and Astronomy, University of Sussex, Brighton; United Kingdom.\\
% ...
$^{158}$Raymond and Beverly Sackler School of Physics and Astronomy, Tel Aviv University, Tel Aviv; Israel.\\
% ...
$^{164}$Department of Physics, University of Toronto, Toronto ON; Canada.\\
% ...
$^{a}$ Also at Borough of Manhattan Community College, City University of New York, NY; United States of America.\\
$^{b}$ Also at Centre for High Performance Computing, CSIR Campus, Rosebank, Cape Town; South Africa.\\
$^{c}$ Also at CERN, Geneva; Switzerland.\\
$^{d}$ Also at CPPM, Aix-Marseille Universit\'e, CNRS/IN2P3, Marseille; France.\\
$^{e}$ Also at D\'epartement de Physique Nucl\'eaire et Corpusculaire, Universit\'e de Gen\`eve, Gen\`eve; Switzerland.\\
% ...
$^{*}$ Deceased

\end{flushleft}
